\documentclass[12pt]{article}
\usepackage{fullpage,amsmath,amssymb,graphicx,slashed,xcolor,mathrsfs,putex}
\usepackage[bookmarks=false]{hyperref}
\newcommand \dd[1]{\textrm{d}#1}

\begin{document}

\preprint{COLO-HEP-591 \\ PUPT-2506 \\Imperial/TP/2016/CR/02}

\institution{CU}{${}^1$Department of Physics, 390 UCB, University of Colorado, Boulder, CO 80309, USA \cr
Center for the Theory of Quantum Matter, University of Colorado, Boulder, CO 80309, USA}
\institution{PU}{${}^2$Joseph Henry Laboratories, Princeton University, Princeton, NJ 08544, USA}
\institution{IMP}{${}^3$Blackett Laboratory, Imperial College London, SW7 2AZ, UK}

\title{Gapped Fermions in Top-down Holographic Superconductors}

\authors{Oliver DeWolfe,${}^\CU$ Steven S.~Gubser,${}^\PU$ Oscar Henriksson,${}^\CU$ and \\ Christopher Rosen${}^\IMP$}

\abstract{We use holography to compute spectral functions of certain fermionic operators in three different finite-density, zero-temperature states of ABJM theory with a broken $U(1)$ symmetry. In each of the three states, dual to previously studied domain wall solutions of four-dimensional gauged supergravity, we find that the fermionic operators have gapped spectra. In one case the gap can be traced to the small charge of the fermions, while in the other cases it is due to a particular interaction that mixes particles and holes.}

\date{September 2016}

\maketitle

\newpage

\section{Overview}
\label{sec:Overview}
When  viewed in the context of the AdS/CFT correspondence, some classical solutions to gauged supergravity theories take on a new life. What was once a novel or even esoteric solution to the supergravity equations of motion may, in a holographic light, serve as a guide to our understanding of the possible phases of strongly coupled matter. 

If such supergravity solutions have no Hawking temperature, their field theory duals are at zero temperature as well, and thus offer a holographic candidate for a ground state of the dual field theory. Understanding which ground states strongly coupled matter can achieve is a central theme of contemporary physics research. It follows that controllable computational frameworks such as holography, which are capable of constructing and probing such states, are highly desirable. 

To exploit the full utility of the holographic methods in a ``controllable" way, it is pragmatic to focus one's attention on the ten- or eleven-dimensional supergravity (SUGRA) theories which provide the low energy limit of the known string theories or M-theory (respectively). The primary advantage, which is substantial, is that in doing so one has access to the full power of the holographic dictionary which relates supergravity modes to operators in known, consistent, and typically well-studied field theories. Field theory results extracted via holography in this way are sometimes classified as ``top-down", to distinguish them from ``bottom-up" models in which a phenomenological gravitational action is concocted and little is known about the presumptive dual field theory. 

In practice, the multitude of dynamical degrees of freedom in these ten- and eleven-dimensional SUGRA theories presents a formidable challenge when it comes time to solve the gravitational equations of motion. To partly circumvent this challenge, it is highly beneficial to identify interesting sectors of the SUGRA that survive consistent Kaluza-Klein truncations to lower dimensional SUGRA theories. By definition, any solution to such a consistent truncation can be oxidized to a solution of the higher dimensional parent theory, and thus the benefits of the top-down embedding are also realized in the lower dimensional theory as well.  A crucial feature  of the consistent truncation approach is that the Kaluza-Klein truncation does not introduce new approximations beyond the ones already present in ten- or eleven-dimensional supergravity; in other words, the oxidized solutions are exact in the original parent supergravity theory.  Likewise, fluctuations around the backgrounds as described by the consistently 
truncated theory can be oxidized exactly to the parent supergravity theory.

In this work we will focus on some properties of solutions to the gauged SUGRA obtained by consistently truncating eleven dimensional SUGRA on a seven-sphere to the lowest lying modes. This is the maximally supersymmetric  gauged SUGRA in four dimensions, which has gauge group $SO(8)$ and about which we will have much more to say in section \ref{sec:SUGRA}. Independent of the details, it is important to note that such truncated theories are often {\it still} too unwieldy to confront head on: the $D=4$ $\mathcal{N}=8$ theory we study here, for example, contains 35 + 35 (pseudo)scalar degrees of freedom.

To avoid the unpleasant technical complications involved in the construction of classical solutions with  large numbers of interacting bosonic fields, a further truncation to singlets under some subgroup $H$ of the SUGRA gauge group $G$ can prove invaluable. In the present work, we have $G = SO(8)$ and thus we will be interested in $H\subset SO(8)$-invariant sectors of the theory that consist of a handful of bosonic fields which nevertheless admit non-trivial solutions holographically dual to interesting phases of strongly coupled matter.

To understand which $H$-invariant sectors may yield interesting results, it is necessary to first identify the broad stroke features of the dual phase of matter one wishes to investigate. In this work, we will be interested in strongly-coupled matter at finite density. From the perspective of the field theory, a finite density can be achieved by turning on a chemical potential $\mu$ for fields carrying charge under some global symmetry current $J^\mu$. An obvious example might be that of a conserved $R$-symmetry current, in which case the field theory Lagrangian is modified by a term 
\begin{equation}
\Delta\mathcal{L}_{\mathrm{QFT}} = \mu J_R^t.
\end{equation}
 For simplicity we focus here on the addition of a chemical potential for a single $U(1)$ factor of the Cartan subalgebra of the (in principle) non-Abelian $R$-symmetry. 
 
 By way of the gauge/gravity dictionary, this global $U(1)$ current is translated into the gravitational language to a bulk $U(1)$ gauge field. So, to study a holographic phase of matter at finite density, a crucial ingredient is a SUGRA truncation which retains at least a single Abelian gauge field. Generically, we will thus focus on consistent truncations of $D=4$ $\mathcal{N}=8$ gauged supergravity that we can write in the form $H\times U(1)\subset SO(8)$ where the bosonic fields of the truncated theory are all invariant under $H$. 

The form of the truncations we are considering clearly leaves open the interesting possibility of including bulk matter which is charged under the $U(1)$ outside of $H$. Background solutions to the SUGRA equations of motion which contain non-trivial profiles for such charged bulk matter are holographically dual to states in which a source for (or expectation value of) an operator carrying global $U(1)$ charge is turned on. Such backgrounds thus correspond to phases which break the global $U(1)$ explicitly or spontaneously (respectively). Geometries where $U(1)$ is broken spontaneously are often referred to as holographic superconductors in the literature, and were studied first in bottom-up constructions at nonzero temperature \cite{Gubser:2008px, Hartnoll:2008vx, Hartnoll:2008kx, Gubser:2008pf} and at zero temperature \cite{Gubser:2009cg, Horowitz:2009ij}. We will be interested in two particular top-down zero-temperature constructions of this type, both of which appear as flows from the maximally symmetric $AdS_4$ vacuum in the UV to a distinct $AdS_4$ region in the IR, characterized by a nontrivial extremum of the scalar potential. The first was originally found as the solution to a compactification of 11D supergravity on a generic Sasaki-Einstein manifold \cite{Gauntlett:2009dn,Gubser:2009gp,Gauntlett:2009bh}, and was later embedded in the $H = SU(4)^-$ truncation of four-dimensional gauged supergravity \cite{Bobev:2010ib}, while the latter was constructed in an $H = SO(3) \times SO(3)$ truncation \cite{Bobev:2011rv}. The $SU(4)^-$ case involves only spontaneous breaking of $U(1)$ and hence is a true superconductor, while in the $SO(3) \times SO(3)$ case the $U(1)$ is explicitly broken. These domain wall geometries represent holographic candidates for finite density, zero-temperature ground states of the dual ABJM theory with a broken $U(1)$ symmetry.

Given such a solution, an immediate question is how to best characterize the holographically dual phase of matter. While some sources, expectation values, and even thermodynamic properties can typically be extracted from the background solution with minimal effort, more detailed information can often be obtained by studying the linear response of the solution to various perturbations. Included in this information are field theory conductivities, viscosities, and various spectral functions that can be related to two-point functions by an assortment of Kubo relations. 

In this work, we continue a line of inquiry \cite{DeWolfe:2012uv,DeWolfe:2013uba,DeWolfe:2014ifa} that is centered on top-down fermionic response in strongly coupled phases of matter. The primary objects of interest in these studies are Green's functions of fermionic operators in the dual field theory. From such correlation functions, one can construct fermionic spectral functions which in turn provide useful  data such as the existence, dispersion, and location of fermionic excitations in the phases of interest. 

In the standard BCS theory of superconductivity, the Fermi surface in the normal state of a superconductor is unstable to the formation of Cooper pairs of fermions below the critical temperature. When these Cooper pairs condense in the superconducting phase, an effective interaction arises which mixes particle and hole excitations, simultaneously destroying the Fermi surface and gapping the fermionic excitation spectrum. Both the gap and the dispersion relation governing the fermionic excitations of the superconducting phase are visible in Angle Resolved PhotoEmission Spectroscopy (ARPES) experiments.

It is natural to wonder whether or not the fermionic excitation spectrum is similarly gapped in superconducting phases of holographic matter.   Bottom-up ``probe" fermions (bulk fermions with constant charge and mass chosen arbitrarily) were studied in the top-down $SU(4)^-$ background in \cite{Gubser:2009dt}, where it was noticed that due to the restored Lorentz invariance of the IR fixed point, the dispersion relations of fermionic excitations have a ``light-cone" structure with unstable modes inside the light-cone and stable modes outside. Moreover, in contrast to the BCS expectation, the fermionic excitations were not necessarily gapped in the superconducting phase, and it was shown that the greater the fermion's charge, the more bands of gapless, stable excitations exist. 

Faulkner et al.~\cite{Faulkner:2009am} studied bottom-up fermions in bottom-up holographic superconductors, introducing a ``Majorana" coupling of a charged fermion $\psi$ to itself along with a scalar $\phi$ of twice the charge,
\begin{equation}\label{MajoranaTerm}
	e^{-1}{\cal L} = \phi^\dagger \,  \psi^T C (\eta + \eta_5 \Gamma_5) \psi + {\rm h.c.} \,,
\end{equation}
 which has a structure reminiscent of the coupling of a Cooper pair $\psi\psi$ to the condensate $\phi$ in an effective BCS Lagrangian. It was shown in \cite{Faulkner:2009am} that such an interaction would generically introduce a gap to the band of fermionic excitations as long as the chirality matrix $\Gamma_5$ was present. The effect of this chiral coupling is to mix ``particle" and ``hole" states, as we review in section~\ref{sec:SO3xSO3}. In \cite{DeWolfe:2015kma} the structure of the top-down fermionic couplings in the $SO(3) \times SO(3)$ solution was described, and a simplified non-chiral fermionic mixing matrix was studied. In this background, similar ``Majorana" interactions occur, although with a charged fermion coupled to a neutral fermion (suggestive of a charged-neutral ``Cooper pair") and it was shown that the non-chiral mixing introduces gaps into some, but not all, bands that were present for probe fermions.

In this work, we present the full top-down fermionic interactions for spin-1/2 fields that do not mix with the gravitino, for both the $SU(4)^-$ and $SO(3) \times SO(3)$ backgrounds. Calculating fermionic spectral functions, we find that both holographic phases have fully gapped fermionic degrees of freedom, though for different reasons. In the $SU(4)^-$ background, the fermion charge is small and probe fermions of this charge already have no bands of stable modes; the full top-down interactions do not change this fact. Moreover, in this background the fermion cannot form the Majorana couplings analogous to Cooper pairing, as both the $SU(4)^-$ group theory and the large charge of the scalar condensate forbid it. In the $SO(3) \times SO(3)$ background, on the other hand, it was known from \cite{DeWolfe:2015kma} that the charge is large enough for probe fermions to be ungapped and if a gap appears it will be due to interactions. We indeed find the top-down couplings between charged and neutral fermions, which include a chirality matrix as advocated in \cite{Faulkner:2009am}, fully gap the fermionic modes.

The structure of this paper is as follows. In section~\ref{sec:SUGRA}, we describe relevant aspects of four-dimensional gauged supergravity. In section~\ref{sec:SU4}, we review the $SU(4)^-$ solution, determine the top-down Dirac equations, and calculate the holographic Green's functions and associated spectral functions, while section~\ref{sec:SO3xSO3} does the same for the $SO(3) \times SO(3)$ solutions. We discuss lessons for strong coupled field theories in section~\ref{sec:lessons}.


\section{4D $\mathcal{N}=8$ gauged supergravity and its Dual}\label{sec:SUGRA}
The $D=4$ $\mathcal{N}=8$ supersymmetric gauged supergravity theory \cite{deWit:1978sh,deWit:1982ig} is the consistent truncation of eleven-dimensional supergravity compactified on a seven-sphere to retain only the supermultiplet of the four-dimensional graviton.
It is holographically dual to the large-$N$ limit of the superconformal theory on a stack of $N$ M2-branes, which may be characterized as ABJM theory \cite{Aharony:2008ug}, a $2+1$ dimensional $U(N)_k\times U(N)_{-k}$ Chern-Simons gauge theory with bifundamental matter, at Chern-Simons level $k=1$.

The bosonic degrees of freedom of the SUGRA are the vierbein, $28$ gauge fields in the adjoint of the gauge group $SO(8)$, and 70 real scalars; the fermions are 8 Majorana gravitini and $56$ Majorana spinors. The scalars parametrize the coset $E_{7(7)}/SU(8)$ as a $56$-bein, which can be written
\begin{equation}\label{eq:V}
\mathcal{V} = 
\left( 
\begin{array}{cc}
u_{ij}^{\phantom{ij}IJ} & v_{ijKL}  \\
v^{klIJ} & u^{kl}_{\phantom{kl}KL} \\
\end{array}
\right) = 
\exp{\left( 
\begin{array}{cc}
0 & \Sigma_{IJKL}  \\
\Sigma^{IJKL} & 0 \\
\end{array}
\right)} \,,
\end{equation}
where raising/lowering indices effects complex conjugation, and the  $\Sigma^{IJKL}\equiv \Sigma^*_{IJKL}$ obey the self-duality relation 
\begin{equation}
\Sigma_{IJKL} = \frac{1}{24}\epsilon_{IJKLMNPQ}\,\Sigma^{MNPQ}.
\end{equation}
The second equality of (\ref{eq:V}) represents a ``unitary" gauge-fixing of the internal $SU(8)$ symmetry, removing the distinction between $SO(8)$ index pairs $[IJ]$ and $SU(8)$ pairs $[ij]$ and allowing one to associate definite $SO(8)$ representations to all the fields: the scalars split into a ${\bf 35}_{\rm v}$ of parity even scalars and a ${\bf 35}_{\rm c}$ of parity odd pseudoscalars (corresponding to the real and imaginary parts of $\Sigma$ respectively), the gravitini are in the ${\bf 8}_{\rm s}$, and the spin-1/2 fields are in the ${\bf 56}_{\rm s}$.

\subsection{The Bosonic Lagrangian}
The bosonic sector of the gauged $\mathcal{N}=8$ theory in four dimensions can be written \cite{deWit:1982ig}
\begin{equation}\label{eq:bSUGRA}
2\kappa^2e^{-1}\mathcal{L} = R - \frac{1}{48}\mathcal{A}_\mu^{ijkl}\mathcal{A}^\mu_{ijkl}-\frac{1}{4}\left[F^+_{\mu\nu IJ}\left( 2 S^{IJ,KL}-\delta^{IJ}_{KL}\right)F^{+\mu\nu}_{KL}+\mathrm{h.c.} \right]-2\mathcal{P} \,,
\end{equation}
where the scalar kinetic tensor $\mathcal{A}_{ijkl}$ is defined through
\begin{equation}\label{eq:inv}
D_\mu \mathcal{V}\cdot\mathcal{V}^{-1} \equiv -\frac{1}{2\sqrt{2}}
\left( 
\begin{array}{cc}
0 & \mathcal{A}_{\mu}^{ijkl}  \\
\mathcal{A}_{\mu \,mnpq}  & 0 \\
\end{array}
\right) \,,
\end{equation}
with the derivative being covariant with respect to both $SO(8)$ and $SU(8)$ indices; for example, for a field $\Phi^I_i$ transforming in the fundamental of both $SO(8)$ and $SU(8)$ we have
\begin{equation}
\label{CovariantDeriv}
D_\mu\Phi^I_i = \nabla_\mu\Phi^I_i - \frac{1}{2}\mathcal{B}_{\mu\phantom{i}i}^{\phantom{\mu}j} \, \Phi_j^I-gA_\mu^{IJ}\,\Phi^J_i \,.
\end{equation}
Note that here we have not yet fixed unitary gauge and so $SU(8)$ and $SO(8)$ indices are distinct.
The definition (\ref{eq:inv}) also implicitly fixes the composite $SU(8)$ connection $\mathcal{B}_{\mu\phantom{i}j}^{\phantom{\mu}i}$:
\begin{equation}\label{eq:B}
\mathcal{B}_{\mu\phantom{i}j}^{\phantom{\mu}i}  = \frac{2}{3}\Big( u^{ik}_{\phantom{ik}LM}\mathscr{D}_\mu u_{jk}^{\phantom{jk}LM}-v^{ikLM}\mathscr{D}_\mu v_{jkLM}\Big) \,,
\end{equation}
where $\mathscr{D}_\mu$ is covariant only with respect to the $SO(8)$ that acts on $I,J$, and ignores $SU(8)$ indices $i,j$. 

 The gauge fields have non-abelian field strengths of the standard form,
$F_{\mu\nu}^{IJ} = 2\partial_{[\mu}A_{\nu]}^{IJ}-2gA_{[\mu}^{IK}A_{\nu]}^{KJ}$
with $F^+$ the (imaginary) self-dual part of the field strength; these
couple to the scalars in their kinetic terms via the $S$-tensor defined as
\begin{equation}
\left( u^{ij}_{\phantom{ij}IJ}+v^{ijIJ}\right) S^{IJ,KL} = u^{ij}_{\phantom{ij}KL} \,.
\end{equation}
Lastly, the scalar potential is given by
\begin{equation}\label{eq:spot}
\mathcal{P} = -g^2\left( \frac{3}{4}|A_1^{ij}|^2 - \frac{1}{24}|A_{2i}^{\phantom{2i}jkl}|^2\right) \,.
\end{equation}
where the scalar-dependent tensors $A_1$ and $A_2$ are defined in terms of the $SU(8)$ covariant $T$-tensor,
\begin{eqnarray}
 A_1^{ij}\equiv\frac{4}{21}T_k^{\phantom{k}ikj} \,, \quad A_{2i}^{\phantom{2i}jkl}\equiv-\frac{4}{3}T_i^{\phantom{i}[jkl]} \,, \quad
 T_i^{\phantom{i} jkl}\equiv(u^{kl}_{\ \ IJ}+v^{klIJ})(u_{im}^{\ \ JK}u^{jm}_{\ \ KI}-v_{imJK}v^{jmKI}) \,.
\end{eqnarray}
The supergravity solutions we discuss in this paper are contained within particular truncations of 4D ${\cal N}=8$ gauged supergravity, in each case involving the $H$-invariant fields in a decomposition $H \times U(1) \subset SO(8)$, with $H = SU(4)$ or $H = SO(3) \times SO(3)$. The bosonic sectors are fairly similar for both resulting truncations; they consist of the metric, one gauge field generating the $U(1)$ gauge group, and at least one scalar field charged under that $U(1)$. We will describe them in 
sections \ref{sec:SU4} and \ref{sec:SO3xSO3}, respectively. 

\subsection{The Spin-1/2 Lagrangian}
The objective of this paper is to compute Green's functions for spin-1/2 operators in finite-density states of ABJM theory. This will be done holographically by solving top-down Dirac equations derived from the quadratic fermion part of the supergravity Lagrangian in the corresponding domain wall geometries. Unlike the bosonic fields in the background geometry, we will not restrict our fermions to being $H$-invariant, but will consider a general spin-1/2 field in the 4D ${\cal N}=8$ gauged supergravity Lagrangian. As a further simplification, however, we will consider only spin-1/2 modes that decouple from the gravitini, as we explain in more detail shortly.

The terms in the $\mathcal{N}
=8$ gauged SUGRA Lagrangian quadratic in spin-1/2 fields are \cite{deWit:1982ig}:
\begin{align}
e^{-1}\mathcal{L}_{\rm \bar\chi \chi} = \frac{i}{12}\Big(\bar{\chi}^{ijk}\Gamma^\mu D_\mu \chi_{ijk}-\bar{\chi}^{ijk}\Gamma^\mu\overleftarrow{D}_\mu \chi_{ijk}\Big)&-\frac{1}{2}\Big(F^+_{\mu\nu IJ}S^{IJ,KL}O^{+\mu\nu KL} + \mathrm{h.c.}\Big)\nonumber\\
&+g\frac{\sqrt{2}}{144}\Big( \epsilon^{ijklmnpq}A_{2lmn}^r\bar{\chi}_{ijk}\chi_{pqr}+\mathrm{h.c.}\Big)\,, \label{eq:Lf}
\end{align}
where the fermion tensor $O^+$ is defined through
\begin{equation}
u^{ij}_{\phantom{ij}IJ}O^{+\mu\nu IJ} = \frac{\sqrt{2}}{288}\epsilon^{ijklmnpq}\bar{\chi}_{klm}\Gamma^{\mu\nu}\chi_{npq} \,,
\end{equation}
and the derivative $D_\mu$ contains the spin connection and the $SU(8)$ connection,
\begin{equation}
D_\mu\chi_{ijk} = 
\nabla_\mu \chi_{ijk}
-\frac{1}{2}\mathcal{B}_\mu^l\,_i\chi_{ljk}-\frac{1}{2}\mathcal{B}_\mu^l\,_j\chi_{ilk}-\frac{1}{2}\mathcal{B}_\mu^l\,_k\chi_{ijl} \,.
\end{equation}
For a summary of our spinor conventions, see appendix \ref{sec:spincon}.
The fermions $\chi_{ijk}$ are totally antisymmetric in $i,j,k$ and are Weyl spinors, with raised/lowered indices corresponding to the two different chiralities \cite{deWit:1978sh}; we may write them as chiral projections of a Majorana spinor $\chi^{ijk}_M$,
\begin{align}
 \chi^{ijk} \equiv P_R \chi^{ijk}_M \hspace{1cm} \text{and} \hspace{1cm} \chi_{ijk} \equiv P_L \chi^{ijk}_M \,,
\end{align}
where $P_L, P_R \equiv (1 \mp \Gamma_5)/2$;
 in a basis where Majorana spinors are real, $\Gamma_5$ is imaginary, and raising/lowering indices again becomes complex conjugation. It is convenient for us to write the Lagrangian in terms of these Majorana spinors, with explicit projection operators. In this case the up/down index structure previously used to distinguish complex representations is broken, since we write all Majorana spinors with indices up, and so we show conjugation explicitly with  a $*$. With this switch in notation and some further processing, our fermion Lagrangian becomes
\begin{multline}\label{FermiLagrangian}
 e^{-1}\mathcal{L}_{\rm \bar\chi \chi} = \frac{i}{12} \bar{\chi}_M^{ijk}\Gamma^\mu\nabla_\mu \chi_M^{ijk} - {i \over 16}\bar{\chi}_M^{ijk}\Gamma^\mu\left(\mathcal{B}_{\mu\phantom{l}i}^{\phantom{\mu}l} + (\mathcal{B}_{\mu\phantom{l}i}^{\phantom{\mu}l})^* - \Gamma_5 (\mathcal{B}_{\mu\phantom{l}i}^{\phantom{\mu}l} - (\mathcal{B}_{\mu\phantom{l}i}^{\phantom{\mu}l})^*)
\right) \chi_M^{ljk} 
\\- {\sqrt{2} \over 576} \epsilon_{ijklmnpq} \bar\chi_M^{klm} \Gamma^{\mu\nu} \left[ F^+_{\mu\nu IJ}S^{IJ,KL} ((u^{-1})_{KL}^{\phantom{KL}ij})^*  P_L + F^-_{\mu\nu IJ}  ({S}^{IJ,KL})^* (u^{-1})_{KL}^{\phantom{KL}ij}P_R \right] \chi_M^{npq} 
 \\+ g\frac{\sqrt{2}}{144}  \epsilon_{ijklmnpq} \bar{\chi}_M^{ijk}\left((A_{2r}^{\phantom{2r}lmn})^*P_L+A_{2r}^{\phantom{2r}lmn}P_R\right)\chi_M^{pqr} \,.
\end{multline}
Note that imaginary parts of the scalar tensors come with an extra factor of $\Gamma_5$; in the case where the scalar Ansatz is real (see e.g. \cite{DeWolfe:2014ifa}) these terms vanish. However, in the current truncations of interest $\Sigma_{IJKL}$ will be complex, and these interactions will play a role. 

To reach this point we have dropped gravitino coupling terms; we can determine the cases for which this is valid using the $H \subset SO(8)$ invariance preserved by the backgrounds. Any term in the full Lagrangian is $SO(8)$-invariant, and hence $H$-invariant.
Since the backgrounds are made of fields in the $H$-invariant truncation, in any spinor/gravitino coupling
 \begin{equation}\label{eq:tcoup}
\mathcal{L}_\Psi\chi = \bar{\Psi}_\mu\Gamma^\mu M(\phi)\chi \,,
\end{equation}
the scalar $\phi$ and hence any $M(\phi)$ is $H$-invariant. (An analogous argument can rule out spinor-gravitino couplings due to $H$-invariant Pauli couplings involving $F_{\mu\nu}$.) Thus the coupling can only exist if $\chi$ and $\bar\Psi_\mu$ transform in representations of $H$ whose product contains a singlet; in general this means $\chi$ and $\Psi_\mu$ only couple if they are in the same representation. (Recall that unlike the bosonic fields in the background geometry, the fermions we study need not be part of the $H$-invariant truncation.) Hence, to avoid such couplings, we can simply decompose the ${\bf 56}_{\rm s}$ and ${\bf 8}_{\rm s}$ into $H$-representations, and choose to study the spinors whose $H$-representations are not shared by any gravitino.

\section{The $SU(4)^-$ Flow}\label{sec:SU4}

In this section we study fermionic response in a zero-temperature geometry solving the equations of an Einstein-Maxwell-scalar theory first obtained in compactifications of 11D SUGRA on Sasaki-Einstein manifolds \cite{Gauntlett:2009dn,Gubser:2009gp,Gauntlett:2009bh}, and later embedded in the $H = SU(4)^-$ truncation of 4D ${\cal N}=8$ gauged SUGRA in \cite{Bobev:2010ib}; the fermions we study are associated to the latter embedding of the bulk theory.

\subsection{The $SU(4)^-$ Truncation}
The $SU(4)^-$-invariant sector of maximal gauged supergravity in four dimensions \cite{Bobev:2010ib} is defined as the fields invariant under the $SU(4)\subset SO(8)$ which leaves invariant the four form \,,
\begin{equation}
\mathcal{W}_{23}^{-} = \mathcal{W}_2^{-} + i \mathcal{W}_3^{-}\equiv  \dd z_1\wedge\dd z_2\wedge\dd z_3\wedge \dd\bar{z}_4 \,,
\end{equation}
where the $z_i$ are coordinates on $\mathbb{C}^4$. The sector contains a neutral pseudoscalar, which we can consistently set to zero, and a charged pseudoscalar which is embedded in the coset representative $\Sigma$ as
\begin{equation}
\Sigma = \frac{i}{2}\mathrm{Im}\left( \omega_3 \mathcal{W}_{23}^{-}\right) = \frac{i}{2}\omega\,\mathcal{W}_3^{-} \,.
\end{equation}
In the final equality the complex scalar $\omega_3 \equiv \omega\,e^{i\alpha}$ has been taken to be real.

The coset representatives are obtained from the exponential of the generators, as per (\ref{eq:V}). To carry out the matrix exponentiation, it is useful to construct the projector 
\begin{equation}
\Pi = \frac{1}{16}\mathcal{W}_3^{-} \cdot \mathcal{W}_3^{-}\qquad \mathrm{where} \qquad (A\cdot B)_{IJKL} \equiv A_{IJMN}B_{MNKL} \,.
\end{equation}
This projector is Hermitian, and squares to itself. Moreover, it satisfies the following useful identities:
\begin{equation}
\Sigma\cdot\Sigma^* = 4\omega^2\,\Pi \qquad \mathrm{and} \qquad \Sigma^*\cdot \Pi = \Sigma^* \,.
\end{equation}
Through explicit computation, one then finds
\begin{equation}
u_{ij}^{\phantom{ij}IJ} = \, \delta_{ij}^{IJ}+(\cosh 2\omega -1)\,\Pi_{ijIJ}\,,\quad \quad
v^{klIJ} =  \,-\frac{i}{4}\sinh 2\omega\,(\mathcal{W}_3^-)_{klIJ} \,.
\end{equation}
The single gauge field in the truncation commutes with $SU(4)^-$ inside $SO(8)$.  In terms of the $z_i$, one can define a Kahler structure on $\mathbb{C}^4$ with Kahler form $J^-$  invariant under the $SU(4)^- \times U(1)$ as:
\begin{equation}
J^- = \frac{i}{2}\Big(\dd z_1\wedge\dd \bar{z}_1 +\dd z_2\wedge\dd \bar{z}_2 +\dd z_3\wedge\dd \bar{z}_3 -\dd z_4\wedge\dd \bar{z}_4 \Big) \,,
\end{equation}
and the $U(1)$ gauge field is then embedded in the $A^{IJ}$ of the $SO(8)$ theory as\footnote{This $\mathcal{A}$ should not be confused with the scalar kinetic tensor defined in (\ref{eq:inv}).}
\begin{equation}\label{eq:ASU4}
A = \frac{1}{\sqrt{2}}\mathcal{A}\, J^-.
\end{equation}
Inserting these ansatze, and defining $\xi \equiv (2/\sqrt{3}) \tanh 2\omega$ to make contact with the conventions of \cite{Gauntlett:2009bh}, we arrive at the Lagrangian governing the $SU(4)^-$ invariant sector of the $\mathcal{N}=8$ theory:
\begin{equation}\label{eq:Lag}
e^{-1}\mathcal{L} = R-\mathcal{F}^2-\frac{3}{2}\frac{|\mathcal{D}\xi|^2}{(1-\frac{3}{4}\xi^2)^2}-\frac{24}{(1-\frac{3}{4}\xi^2)^2}(-1+\xi^2) \,,
\end{equation}
where $\mathcal{F} = \dd \mathcal{A}$. In this section we employ conventions such that $g^2 = 2$ and $G_N = 1/(8\pi)$. The covariant derivative is thus given by $D_\mu\xi = \partial_\mu\xi-4i \mathcal{A}_\mu\xi$, and the scalar has charge 4.

To see this from the group theory point of view, under the $SO(8) \to SU(4)^- \times U(1)$ decomposition, the gauge fields transform as
\begin{equation}
{\bf 28} \to {\bf 15}_0 \oplus {\bf 6}_2 \oplus {\bf 6}_{-2} \oplus{ \bf 1}_0 \,,
\end{equation}
where the ${\bf 1}_0$ is our ${\cal A}$, and the parity-even and parity-odd pseudoscalars decompose as
\begin{equation}
\label{Branching35}
{\bf 35}_{\rm v} \to {\bf 15}_0 \oplus {\bf 10}_2 \oplus {\bf \overline{10}}_{-2} \,, \quad \quad
{\bf 35}_{\rm c} \to  {\bf 20'}_0 \oplus {\bf 6}_2 \oplus {\bf 6}_{-2}\oplus {\bf 1}_4 \oplus {\bf 1}_{-4} \oplus {\bf 1}_{0} \,, 	
\end{equation}
so our charged scalar $\xi$ (or $\omega_3$) is the ${\bf 1}_4$ and its conjugate. 
	
\subsection{$SU(4)^-$-invariant Domain Wall Solutions}\label{sec:PWAdS}

The zero-temperature solution we are interested in corresponds to a flow driven by a relevant deformation from the maximally supersymmetric $AdS_4$ geometry in the UV to the so-called Pope-Warner $AdS_4$ solution \cite{Pope:1984bd} in the IR \cite{Gauntlett:2009dn,Gubser:2009gp,Gauntlett:2009bh}. This deformation does not involve adding a scalar operator to the dual Lagrangian; the relevant deformation is a spatially uniform chemical potential, and the response of the scalar operator is only to acquire an expectation value, so the geometry is a true holographic superconductor with $U(1)$ broken only spontaneously.

The chemical potential breaks Lorentz invariance as well as conformal invariance, but when it leads to a domain wall solution between two $AdS_4$ vacua, full relativistic conformal invariance is recovered in the infrared as an emergent symmetry.  A striking feature is that the speed of light $v_{IR}$ in the infrared is smaller than the speed of light $v_{UV}$ in the ultraviolet---meaning simply that $g_{tt}/g_{xx}$ has different IR and UV limits.  Physically, we can 
think of the ratio $v_{UV}/v_{IR}$ as an 
index of refraction for the holographic state of matter that we are describing.  By rescaling $\vec{x}$, we can change $v_{UV}$ and $v_{IR}$ by the same factor, but the index of refraction remains invariant.  An interesting conjecture \cite{Gubser:2009gp} states (approximately) that the type of deformation we study, based on a chemical potential and flowing to an infrared conformal fixed point, always exists in holographic theories provided there is an associated renormalization group flow preserving Lorentz invariance throughout with the same UV and IR conformal fixed points. 

The $SU(4)^-$ holographic superconductor geometry is encapsulated by the ansatz
\begin{equation}\label{eq:DWansatz}
\dd s^2 = -G(r)e^{-\beta(r)}\dd t^2 + \frac{\dd r^2}{G(r)}+r^2\dd\vec{x}^2, \qquad \mathcal{A} = \phi(r)\, dt, \qquad \mathrm{and} \qquad \xi = \xi(r).
\end{equation}
The maximally supersymmetric $AdS_4$ vacuum has $G=4r^2$ and $\beta = \phi = \xi = 0$, while the PW $AdS_4$ solution has $\xi = \sqrt{2/3}$, corresponding to another extremum of the potential \eno{eq:Lag}, as well as $G=16r^2/3$ and $\beta = \phi = 0$.

To construct the flow between the $AdS_4$ solutions, it is helpful to characterize the spectrum of irrelevant perturbations of the PW solution, as these can be used to integrate away from the solution towards the maximally symmetric solution in the UV.
Linearizing the equations of motion about the PW solution, one finds that there is a scalar mode and a vector mode both with mass $m^2 = 6$ which satisfy the flow criteria. They are holographically dual to scalar and vector operators of the IR conformal field theory with dimension $\Delta = (3+\sqrt{33})/2$ and $\Delta = 4$ respectively. 
The linearized analysis fixes the irrelevant perturbations to be of the form
\begin{equation}
G = \, \frac{16}{3}r^2+\ldots \,, \quad
\beta = \, 4 + \ldots\,, \quad
\phi = \, r^2 + \ldots\,,\quad
\xi = \, \sqrt{\frac{2}{3}}+\mathcal{J}r^{\frac{1}{2}(-3 +\sqrt{33})}+\ldots
\end{equation}
Scaling symmetries of the equations of motion have been used to fix the amplitudes of the $\beta$ and $\phi$ perturbations arbitrarily, leaving only a single parameter $\mathcal{J}$ to be tuned such that the desired behavior is obtained in the UV.

In the UV, the scalar $\xi$ provides a $\Delta = 2$ perturbation of the maximally symmetric $AdS_4$. Since we are interested in the case when the UV fixed point is not deformed by a source for the dual scalar operator, the UV behavior of the scalar is required to be of the form $\xi(r\to\infty) \sim \xi_2/r^2 + \ldots$ representing a spontaneously acquired vacuum expectation value for the dual scalar operator.

\begin{figure}
\centering
\includegraphics[scale=0.4]{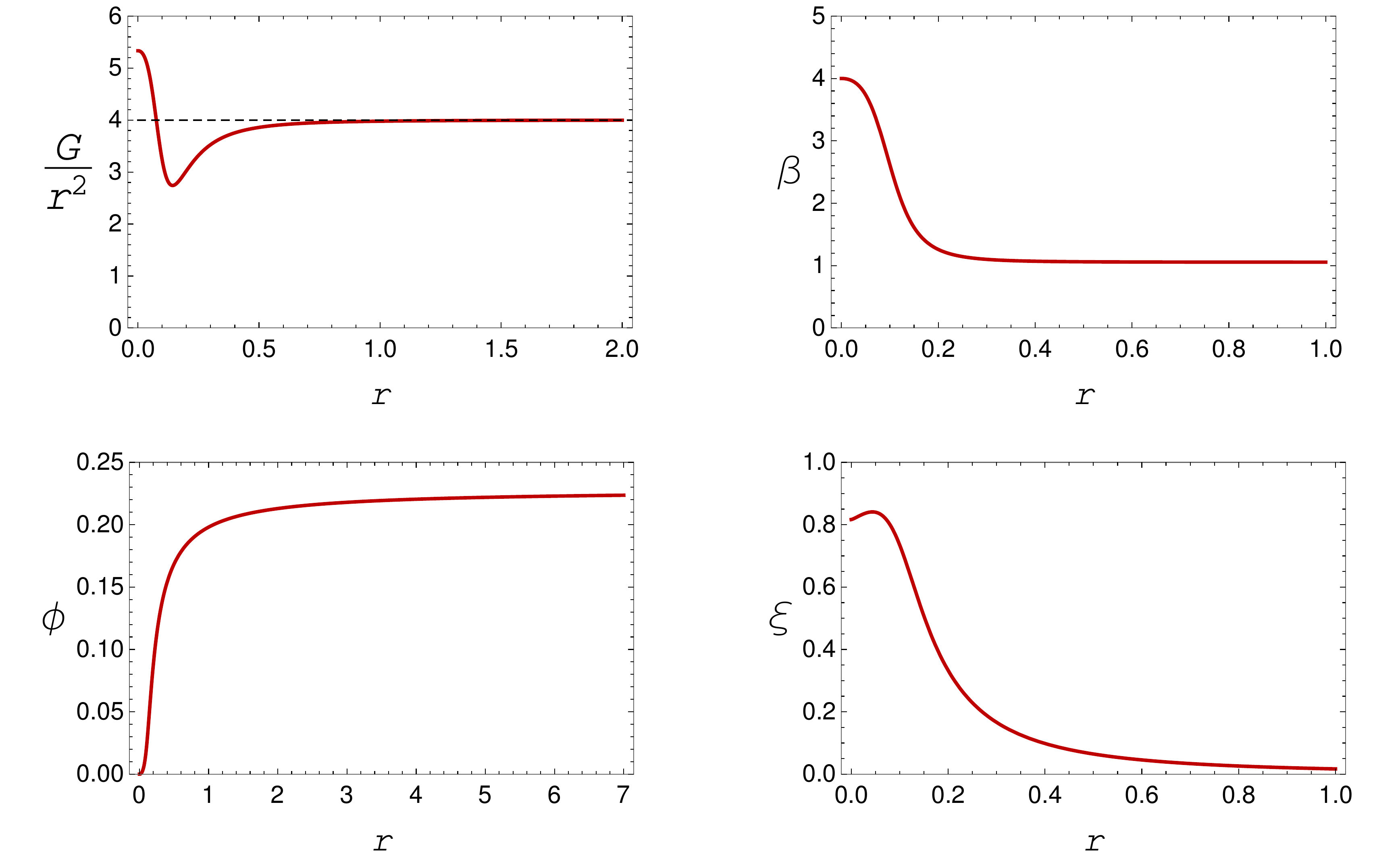}
\caption{The $AdS_4$ to PW flow. The flow is characterized by an index of refraction $n \approx 3.78$ and a scalar vev proportional to $\xi_2/\phi_{\mathrm{UV}}^2\approx0.33$.}
\label{fig:DW}
\end{figure}

The desired solution is readily constructed from a numerical shooting technique, tracing the RG flow upstream to the UV. It appears in figure \ref{fig:DW}.

\subsection{The Fermionic Sector}\label{sec:SU4ferm}
The supersymmetries of the gauged supergravity transform in the ${\bf 8}_s$, which decomposes as  
\begin{equation}\label{eq:embed}
{\bf 8}_s\to {\bf 4}_{-1}\oplus {\bf \bar{4}}_{1}\,,
\end{equation}
 under the $SU(4)^-\times U(1)$. Accordingly no supersymmetries survive the truncation. In fact, since the spin-1/2 fermions decompose as
 \begin{equation}
 {\bf 56}_s\to {\bf 20}_{-1}\oplus {\bf \overline{20}}_{1}\oplus\bf {4}_{3}\oplus {\bf \bar{4}}_{-3}\oplus\bf {4}_{-1}\oplus {\bf \bar{4}}_{1} \,,
  \end{equation}
  there are no singlets in the fermionic sector of the truncation at all---the $SU(4)^-$ invariant theory is entirely bosonic. 
From our perspective this is not a problem, as we are happy to study any ${\cal N}=8$ gauged supergravity spin-1/2 fields, regardless of whether they are in the $SU(4)^-$ truncation. We only wish to avoid gravitino mixing, which we can do as long as we avoid the ${\bf 4}$ representations, and study instead the ${\bf 20}_{-1}$ and its conjugate.

Churning through the various supergravity tensors in \eno{FermiLagrangian}, one arrives at a Lagrangian for the spin-1/2 modes of the form
\begin{equation}\label{eq:gLag}
e^{-1}\mathcal{L}_{\bar{\chi}\chi} = \bar{\chi}\Big(\slashed{\nabla}+\slashed{\mathbb{B}}+\slashed{\mathbb{P}}+\mathbb{M}\Big)\chi \,,
\end{equation}
where the $\chi$ are considered as 56 component vectors whose entries are the non-vanishing $\chi_{ijk}$, $\nabla$ is the derivative covariant on the background geometry, and $\mathbb{B}, \mathbb{P},\mathbb{M}$ are 56 by 56 dimensional matrices describing the gauge, Pauli, and mass couplings, respectively. We are using a schematic ``slashed" notation to indicate the appropriate Lorentz invariant contraction with the $\Gamma^\mu$. 

We can now isolate the Dirac equations for the fields in the ${\bf 20}$. Due to an $SU(4)$-invariance argument analogous to the argument for the gravitino/spin-1/2 coupling in \eno{eq:tcoup}, each member of the ${\bf 20}$ cannot mix with anything but itself. This forbids mixing with its own conjugate (which is in the inequivalent ${\bf \overline{20}}$ representation), ruling out ``Majorana" couplings of the type shown in \eno{MajoranaTerm}. It is helpful to note that $\slashed{\mathbb{B}}$ and  $\slashed{\mathbb{P}}$ commute in this case, and thus the kinetic, gauge, and Pauli terms can be simultaneously diagonalized. In this basis, the decomposition is manifest.

In terms of the $\chi_{ijk}$, a representative of the  ${\bf 20}_{-1}$ can be chosen to be the  combination
\begin{equation}\label{eq:SU4fermion}
\psi = \chi_{368}+\chi_{467}+i (\chi_{358}+\chi_{457}) \,,
\end{equation}
and this or any other fermion in the ${\bf 20}$ can then be seen to satisfy the Dirac equation
\begin{equation}\label{eq:DEQ20}
\left(i\slashed{\nabla}-\frac{4+3\xi^2}{4-3\xi^2}\slashed{\mathcal{A}}+\frac{i}{4}\slashed{\mathcal{F}}-\frac{3\xi^2}{4-3\xi^2}\right)\psi = 0.
\end{equation}
At the UV fixed point, the scalar $\xi$ vanishes and $\psi$ is massless, and thus from the perspective of the ABJM theory, $\psi$ is dual to operators  carrying charge $|q_\psi| = 1$ and having conformal dimension $\Delta = 3/2$. Comparing to the complex scalar $\xi$ of  (\ref{eq:Lag}) with charge $|q_\xi |= 4$, one  finds that the scalar carries four times the $U(1)$ charge of the decoupled fermions. This is another reason why ``Majorana" couplings of the type \eno{MajoranaTerm} are forbidden in this case. 

Along the flow $\xi$ runs from 0 to the IR value $\xi = \sqrt{2/3}$. Thus in the IR theory governed by the PW solution, the supergravity mode $\psi$ behaves as though it carries mass $m_{IR} = 1$.

\subsection{Fermion Response} 

We now wish to solve the equation (\ref{eq:DEQ20}) in the background of Figure \ref{fig:DW}. We use the basis (\ref{eq:gbasis}) for the generators $\Gamma^a$, and to label the four complex components of our spinors we define the projectors \cite{Faulkner:2009wj}
\begin{equation}\label{Projectors}
 \tilde{\Pi}_\alpha \equiv {1 \over 2} \left(1 - (-1)^{\alpha}i \Gamma^{\hat{r}} \Gamma^{\hat{t}} \Gamma^{\hat{x}} \right) \,,\quad \quad
P_\pm \equiv {1 \over 2}\left ( 1 \pm i \Gamma^{\hat{r}} \right) \,,
\end{equation}
where $\alpha = 1,2$. One can then write the four components of a bulk spinor $\psi$ as
\begin{equation}
 \psi_{\alpha \pm} \equiv \tilde{\Pi}_\alpha P_\pm \psi \,.
\end{equation}
From a 2+1 dimensional point of view, $\psi_+$ and $\psi_-$ each transform as Dirac spinors, and $\alpha$ labels the two complex components of these spinors. As discussed in \cite{DeWolfe:2014ifa}, supersymmetry fixes $\psi_+$ to be the spinor that asymptotes to a source for the dual fermionic operator. 

It is computationally convenient to ``square" (\ref{eq:DEQ20}) to arrive at second order linear differential equations governing the components of $\psi_+$. We also redefine the spinor as
\begin{equation}
\psi(t,r,x)\to (G\, r^4e^{-\beta})^{-\frac{1}{4}}\psi(r)\,e^{-i(\omega t-k x)}\,,
\end{equation}
where we have exploited the background isometries to set the momentum in the $x$-direction, and the metric factor has been chosen to cancel the spin-connection part of the covariant derivative. In practice, the basis we adopt allows one to focus on either the $\alpha = 1$ or $\alpha = 2$ components independently. Rotational invariance of the background then ensures that $\psi_1(k) = \psi_2(-k)$.

Asymptotically, in the UV the source components behave like
\begin{equation}\label{eq:UVsol}
\psi_+(r)\sim J(\omega, k)+\mathcal{O}(1/r) \,,
\end{equation}
where $J(\omega,k)$ is interpreted holographically as a source for the dual fermionic operator. In the IR, these components obey an equation of the form
\begin{equation}\label{eq:IReq}
\psi_+'' +\frac{2}{r}\psi_+'-\Big(\frac{\tilde{m}(1+\tilde{m})}{r^2}+\frac{L_\mathrm{IR}^2p^2}{r^4}\Big)\psi_+=0 \,,
\end{equation}
where $\tilde{m} = m_\mathrm{IR}L_\mathrm{IR}=L_\mathrm{IR}$ is the dimensionless mass of the fermion in the IR and $L_\mathrm{IR}$ of the AdS radius in the PW solution. We have also introduced
\begin{equation}\label{eq:IRmom}
p^2 \equiv -\frac{\omega^2}{v_{IR}^2}+k^2 \,,
\end{equation}
the Lorentz invariant momentum squared of the mode in the PW background, where $v_\mathrm{IR}$ is the speed of light in the PW solution.

The features of the solution to (\ref{eq:IReq}) depend strongly on the sign of $p^2$. The case of spacelike momentum $p^2>0$ is particularly interesting. This is because for a system consisting of a finite density of fermions, one might expect to find significant spectral weight at zero energy (as measured from the chemical potential) but non-vanishing momentum. In that situation,  (\ref{eq:IReq}) is solved by a component of the form
\begin{equation}\label{eq:IRsol}
\psi_+ = \frac{1}{\sqrt{r}}K_{-\frac{1}{2}-\tilde{m}}\Big( \frac{p\,L_\mathrm{IR}}{r}\Big)\,,
\end{equation}
with $K$ the modified Bessel function of the second kind. The IR Green's function $\mathcal{G}_R(\omega,k)_{ \alpha \beta}$ can be a useful diagnostic to quantify the fermion response. To construct it, note that the Dirac equation (\ref{eq:DEQ20}) implies that
\begin{align}
\psi_- = &\frac{r^2}{L_\mathrm{IR}}v_\mathrm{IR}\left(\frac{1}{k\, v_\mathrm{IR}+\omega }\right)\left(\frac{\tilde{m}}{r}\psi_+ -  \psi_+'\right)\nonumber\\
 = & -\frac{p\,v_\mathrm{IR}}{\sqrt{r}}\left(\frac{1}{k\, v_\mathrm{IR}+\omega }\right)K_{-\frac{1}{2}+\tilde{m}}\Big( \frac{p\,L_\mathrm{IR}}{r}\Big)\,,
\end{align}
where $\psi_-$ is the component of the bulk spinor whose normalizable fall-off encodes the field theory response. Applying the holographic prescription for the dual retarded correlator thus gives
\begin{equation}
\mathcal{G}_R(\omega,k)\,_{ \alpha, \beta = 1}= -\frac{1}{4^{\tilde{m}}}\frac{\Gamma(\frac{1}{2}-\tilde{m})}{\Gamma(\frac{1}{2}+\tilde{m})}\frac{(p\,L_\mathrm{IR})^{2\tilde{m}}}{k\, v_\mathrm{IR}+\omega}p\,v_\mathrm{IR}
\end{equation}
where $\alpha,\beta$ are spinor indices. To construct a rotationally invariant correlator one can trace over the spinor indices to obtain
\begin{align}\label{eq:trGR}
\mathcal{G}_R &\equiv \mathrm{tr}\,\mathcal{G}_R\,_{\alpha\beta} = \mathcal{G}_R(\omega,k)\,_{ 11}+\mathcal{G}_R(\omega,-k)\,_{ 11}\nonumber\\
& = -\frac{1}{2^{2\tilde{m}-1}}\frac{\Gamma(\frac{1}{2}-\tilde{m})}{\Gamma(\frac{1}{2}+\tilde{m})}\frac{(p\,L_\mathrm{IR})^{2\tilde{m}}}{p\,v_\mathrm{IR}}\omega.
\end{align}
In performing the trace, we have exploited the fact that in this system $\mathcal{G}_R(\omega,k)_{22}=\mathcal{G}_R(\omega,-k)_{11}$  as a consequence of  the dual state's isotropy.

The domain wall background of figure \ref{fig:DW} departs fairly quickly from the PW solution which characterizes the IR, and thus one expects that IR Green's functions such as  (\ref{eq:trGR}) characterize the field theory dynamics only for those bulk fermion solutions which are localized very near $r=0$. 

The solution (\ref{eq:IRsol}) is regular as $r\to 0$, and purely real. Its form suggests the interesting possibility of constructing fermion normal modes in the domain wall solution which behave like (\ref{eq:IRsol}) in the IR and asymptote to (\ref{eq:UVsol}) in the UV with $J = 0$ for some choice of $(\omega,k)$. Indeed, such fermion normal modes were observed in various bottom-up holographic models, such as \cite{Gubser:2009dt,Faulkner:2009am,DeWolfe:2015kma}. We now attempt to construct these as linearized perturbations of the $SU(4)^-$ invariant flow.

Solving the bulk Dirac equation (using numerical shooting from the IR to the UV) and scanning over spacelike momenta reveals a null result: we find no fermion normal modes for the fermions in the {\bf 20} or ${\bf \overline{20}}$. In particular, there is no mode at $\omega = 0$, and thus the fermionic spectral function is gapped in this state of the ABJM theory. 

To quantify and better visualize the fermion response one can look to the spectral function of the dual field theory operators, which we define to be 
\begin{equation}\label{eq:spectralFunc}
A(\omega,k) = \frac{i}{2}\mathrm{tr}\left({\bf G}_R-{\bf G}_R^\dagger \right).
\end{equation}
Here ${\bf G}_R$ is the $2\times 2$ matrix of retarded Green's functions for the two-component fermionic operators. To extend the domain of the spectral function to timelike momenta, one must modify the IR boundary condition (\ref{eq:IRsol}) to provide the proper notion of ``ingoing" necessary to reproduce the causal structure of the retarded correlator. The correct prescription is given in \cite{Iqbal:2009fd}, and turns out to be
\begin{equation}
\psi_+ = \left\{ \begin{array}{ll}
 \frac{1}{\sqrt{r}}H^{(1)}_{-\frac{1}{2}-\tilde{m}}\Big( \frac{\sqrt{- p^2}\,L_\mathrm{IR}}{r}\Big)
 & \omega > v_\mathrm{IR}|k|\\
 \frac{1}{\sqrt{r}}H^{(2)}_{-\frac{1}{2}-\tilde{m}}\Big( \frac{\sqrt{-p^2}\,L_\mathrm{IR}}{r}\Big) &\omega < v_\mathrm{IR}|k|
  \end{array} \right.
\end{equation}
 with $H$ the Hankel function of the first or second kind as indicated. 
 
\begin{figure}
\centering
\includegraphics[scale=0.42]{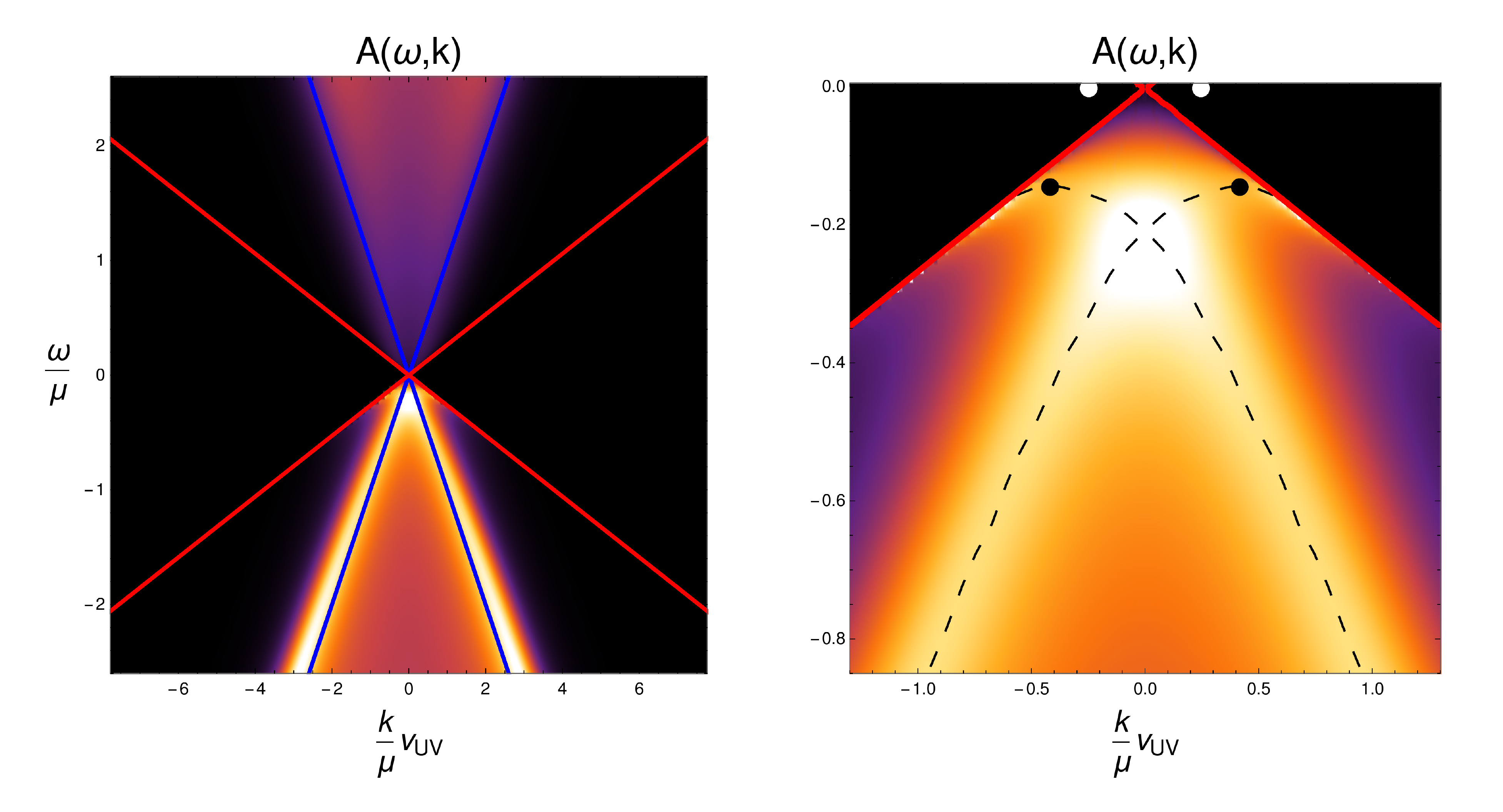}
\caption{Spectral function for fermionic operators in the {\bf 20}. The red lines mark the IR lightcone, while the blue lines show the lightcone of the UV theory. The right figure shows a close-up around the origin for $\omega<0$. Superimposed on the right figure are black dashed lines, showing the lines of maxima of the spectral weight; black dots, marking the point of closest approach to the $\omega=0$ axis ($k^\star$); and white dots, showing the location of the Fermi surface singularities in the normal phase ($k_F$). These special points will be discussed in more detail in section \ref{sec:lessons}.}\label{fig:spec}
\end{figure}

 The spectral function is shown in figure \ref{fig:spec}. Notably, from the leftmost plot one observes that for sufficiently large values of $\omega/\mu$ the spectral weight is confined to the edges of a roughly conical structure in momentum space with slope one in the units of the figure. This is in fact the conformal behavior anticipated from the presence of the maximally symmetric $AdS_4$ in the UV. This can readily be seen from the analytic continuation of (\ref{eq:trGR}) to timelike momenta, replacing the labels ``IR" with ``UV", and evaluating $\tilde{m} =0$. To wit, for $\omega > v_{\mathrm{UV}}|k|$ one obtains
 \begin{equation}\label{eq:UVspectralWeight}
 A(\omega,k) = \frac{2}{v_{\mathrm{UV}}}\frac{\omega}{\sqrt{-p^2}},
 \end{equation}
 where the Lorentz contraction implied by $p^2$ is now understood to be with respect to the maximally symmetric $AdS_4$ metric.  The right plot in figure \ref{fig:spec} shows the spectral function zoomed-in around the origin for $\omega<0$. The spectral function in this region is somewhat diffuse, hence we have added two black dashed lines which trace out the peaks in the spectral weight of the two spinor components of the fermionic operator. These lines clarify that the bands of the two spinor components cross at $k=0$ and reach their turning points at some non-zero $k$; this is reminiscent of a holographic Rashba effect, as was discussed previously in \cite{Herzog:2012kx}.

\begin{figure}
\centering
\includegraphics[scale=0.5]{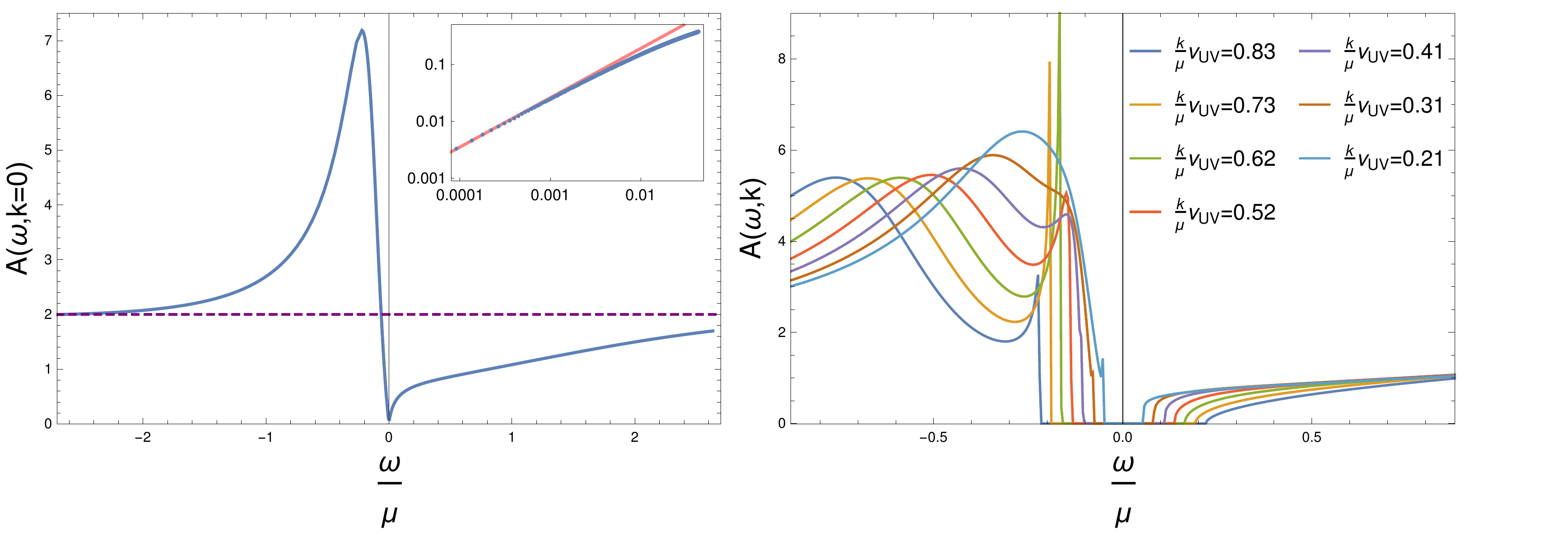}
\caption{Spectral function for fermionic operators in the {\bf 20} as a function of frequency at various momenta. At left, $k=0$ and the dashed purple line shows the maximally symmetric $AdS_4$ result as given by (\ref{eq:UVspectralWeight}). The inset details the falloff at low frequencies, which asymptotes to a power law with exponent $2 \Delta_{IR}-3 = \sqrt{3}/2$ as shown by the pink line. At non-zero momenta (right), the spectral function develops a hard gap. For momenta in the vicinity of $k \approx k^\star$ there is a narrow quasiparticle-like peak just below the gap, as well as a more diffuse hump at larger $|\omega/\mu|$ as dictated by the UV conformal theory. }
\label{fig:SU4_k0}
\end{figure}

To further quantify the properties of any putative  fermionic excitations, it is also helpful to consider the spectral weight along several representative momentum slices. Strictly at $\omega = k = 0$,  the bulk mode decays in the IR as a power law, and one can explicitly show that the spectral weight vanishes at this point. Extending this computation to finite $\omega$ along $k=0$ results in the slice shown in the left plot of figure \ref{fig:SU4_k0}. Most notably, the spectral weight exhibits a ``soft gap", vanishing like a power law as $\omega\to 0$. By studying the properties of the IR Green's functions along this slice, it is straightforward to demonstrate that 
 \begin{equation}
 A(\omega,k=0) \sim \omega^{2\Delta_\mathrm{IR}-3} \qquad \mathrm{for} \qquad \frac{\omega}{\mu}\ll 1 \,,
 \end{equation}
 where $\Delta_\mathrm{IR} = \frac{1}{2}(3+2\tilde{m})$ is the conformal dimension of the fermionic operator in the IR theory.

 The slices along non-zero momenta are rather more interesting. From the right plot of figure \ref{fig:SU4_k0},  one can clearly distinguish the appearance of the hard gap in the spectral weight corresponding to the boundaries of the IR lightcone. As the momentum is increased from zero, the broad peak controlled by the UV fixed point develops a shoulder near the gap, which eventually sharpens into a well defined secondary peak. This secondary peak is present for momenta $k\approx k^\star$ which is the momentum at which the maximum of the arcing spectral weight achieves its closest approach to $\omega = 0$. Accordingly it is natural to associate this secondary peak with a gapped fermionic excitation in the dual ABJM phase of matter. We will have more to say about this excitation and its holographic interpretation in section \ref{sec:lessons}. For now, we note that these spectral functions share similarities with the ``peak-dip-hump" structure observed in various ARPES measurements of the high $T_c$ superconductors. This experimental structure has been argued to be a consequence of many-body interactions in the superconducting phase (eg. \cite{2002cond.mat..9476C}). Similar line shapes were observed holographically in \cite{Chen:2009pt} and \cite{Faulkner:2009am}. The link between these results and the experimentally observed peak-dip-hump is tenuous; in our current case the pattern is likely a consequence of the previously mentioned Rashba-like crossing of two bands, in combination with the sharpening of the peaks as they approach the IR lightcone.

\subsection{Field Theory Operator Matching}
To make contact with the dual field theory, it is necessary to first employ the holographic dictionary to translate the bulk fields involved in our solutions into field theory operators. These operators are distinguished by their quantum numbers---conformal dimensions and charges under various symmetry groups.

The dual superconformal field theory is most commonly written in terms of ABJM theory \cite{Aharony:2008ug}, a Chern-Simons-matter theory which makes a global $SU(4) \times U(1)_b \subset SO(8)$ manifest, while the full $SO(8)$ is present but not apparent in the Lagrangian. However, this $SU(4)$ subgroup and the commuting $U(1)_b$ (associated with monopole charge) are different from the $SU(4)^- \times U(1)$ subgroup relevant to our geometry; the two sets of subgroups are related by a triality transformation.

The supercharges in the ${\bf 8}_{\rm s}$ decompose under $SU(4)^- \times U(1)$ as ${\bf 8}_s\to {\bf 4}_{-1}\oplus {\bf \bar{4}}_{1}$ \eno{eq:embed} but under the $SU(4) \times U(1)_b$ of ABJM theory as ${\bf 8}_s\to {\bf 6}_0 \oplus {\bf 1}_2 \oplus {\bf 1}_{-2}$. This latter decomposition aligns with the isometries  of the moduli space for a stack of M2-branes probing a $\mathbb{C}^4/\mathbb{Z}_k$ singularity \cite{Aharony:2008ug}. The former branching, on the other hand, corresponds to the decomposition of the supersymmetries when the sign of the M2-brane charge is reversed. Reversing the sign of the M2-brane charge is realized in the eleven dimensional SUGRA as a ``skew-whiffed" solution in which the four-form flux has opposite sign (or, equivalently, the orientation of the $S^7$ is reversed). Indeed, when the PW solution is oxidized to eleven dimensions, the solution is of this skew-whiffed form \cite{Pope:1984bd,Bobev:2010ib}. The flows constructed in section \ref{sec:PWAdS} thus connect the PW solution to a skew-whiffed $AdS_4$ in the 
UV. For Chern-Simons level $k=1$, the skew-whiffed $AdS_4\times S^7$ still preserves maximal supersymmetry, and the holographic dual remains the ABJM theory. This is the case relevant for the holographic interpretation of our supergravity results. The skew-whiffing is then realized from the field theory perspective as a triality rotation on the operator spectrum \cite{Forcella:2011pp}, as might be anticipated from the various decompositions of the global symmetries we have considered.

Because the two $SU(4)$ groups do not commute, representations of $SU(4)^-$ do not fill out complete representations of the ABJM $SU(4)$. Instead of presenting dual operators in the ABJM language, we will instead use a simplified notation with manifest $SO(8)$ invariance, which we can think of as a generalization of the theory living on a single M2-brane: we will combine 8 field theory scalars in the ${\bf 8}_{\rm v}$ into complex combinations $Z_i$, $i= 1, 2, 3, 4$, and 8 field theory Majorana spinors in the ${\bf 8}_{\rm c}$ into complex combinations $\Lambda_i$, $i = 1, 2, 3, 4$.

In this notation, the operator dual to the complex scalar turned on in the background is the $\Delta=2$ fermion bilinear,
\begin{equation}
\xi \quad \leftrightarrow \quad \Lambda_1 \Lambda_1 \,.
\end{equation}
The gauge field \eno{eq:ASU4} corresponds to the chemical potentials for the four Cartan generators of $SO(8)$ being identified as 
\begin{equation}
\label{SU4Mu}
	\mu_a = \mu_b = \mu_c = - \mu_d \,.
\end{equation}
The fermionic supergravity fields are then dual to scalar/fermion composite operators with dimension $\Delta = 3/2$ of the form $Z \Lambda$. 
The mode \eno{eq:SU4fermion} is the linear combination
\begin{equation}
	\psi \quad \leftrightarrow \quad \bar{Z}_3 \Lambda_2 - \bar{Z}_4 \Lambda_4\,.
\end{equation}

\section{The $H = SO(3)\times SO(3)$ Flow}\label{sec:SO3xSO3}
 We next turn our attention to a similar pair of domain wall geometries found within an $SO(3)\times SO(3)$ invariant truncation of the gauged SUGRA \cite{Bobev:2011rv}. As before, these backgrounds are holographically dual to zero temperature phases of ABJM theory with a broken $U(1)$ global symmetry, though in this case it is explicitly as well as spontaneously broken. Again we will discover a gapped fermion excitation spectrum in these states. This time, however, the gapping mechanism relies on a special type of fermion coupling, similar to the ``Majorana coupling" previously studied in the bottom-up construction of \cite{Faulkner:2009am}.

\subsection{The $SO(3) \times SO(3)$ Truncation and Domain Wall Solutions}
To truncate the full supergravity to the $SO(3)\times SO(3)$ invariant sector, we make the following ansatz for the scalar tensor \cite{Bobev:2011rv}:
\begin{equation}\label{eq:sanz}
\Sigma_{IJKL}=\frac{\lambda}{2\sqrt{2}}\left[ \cos \alpha \Big(\mathcal{Y}^+_{IJKL}+i\, \mathcal{Y}^-_{IJKL} \Big) - \sin \alpha \Big(\mathcal{Z}^+_{IJKL}-i\, \mathcal{Z}^-_{IJKL} \Big) \right]\,.
\end{equation} 
Here, $\lambda$ and $\alpha$ are four-dimensional scalars, and $\mathcal{Y}^{\pm}$ and $\mathcal{Z}^{\pm}$ are self dual (+) and anti-self dual ($-$) invariant four-forms on the scalar manifold:
\begin{eqnarray}
\mathcal{Y}^+ = & \dd x_3\wedge\dd x_4\wedge\dd x_5\wedge\dd x_1 +\dd x_2\wedge\dd x_6\wedge\dd x_7\wedge\dd x_8 \,, \nonumber\\
\mathcal{Y}^- = & \dd x_3\wedge\dd x_4\wedge\dd x_5\wedge\dd x_2 +\dd x_1\wedge\dd x_6\wedge\dd x_7\wedge\dd x_8 \,, \nonumber\\
\mathcal{Z}^- = &\dd x_3\wedge\dd x_4\wedge\dd x_5\wedge\dd x_1 -\dd x_2\wedge\dd x_6\wedge\dd x_7\wedge\dd x_8 \,, \nonumber\\
 \mathcal{Z}^+= &\dd x_3\wedge\dd x_4\wedge\dd x_5\wedge\dd x_2 -\dd x_1\wedge\dd x_6\wedge\dd x_7\wedge\dd x_8 \,.
\end{eqnarray}
Here the $x_i$ are coordinates on the $\mathbb{R}^8$ of $SO(8)$. In this language,  $(x_3,x_4,x_5)$ and  $(x_6,x_7,x_8)$ transform as the fundamental representation under different $SO(3)$ factors in $H$. Evaluating the Lagrangian (\ref{eq:bSUGRA}) in the $SO(3)\times SO(3)$ invariant truncation gives 
\begin{equation}\label{eq:Lbose}
e^{-1}\mathcal{L} = \frac{1}{2}R-\frac{1}{4}F_{\mu\nu}F^{\mu\nu}-\partial_\mu\lambda\partial^\mu\lambda-\frac{\sinh^2(2\lambda)}{4}\left(\partial_\mu\alpha-g\mathcal{A}_\mu\right)\left(\partial^\mu\alpha-g\mathcal{A}^\mu\right)-\mathcal{P} \,,
\end{equation}
where $\kappa^2$ has now been set to one,  and the  remaining $U(1)$ is embedded in the $A^{IJ}$ like 
\begin{equation}\label{eq:ASO32}
A = \mathcal{A}\,\dd x_1\wedge\dd x_2.
\end{equation}
The scalar potential is
\begin{equation}\label{eq:scalarV}
\mathcal{P} = \frac{g^2}{2}\left(s^4-8s^2-12 \right) \qquad \mathrm{with} \qquad s \equiv \sinh\lambda \, ,
\end{equation}
and it has critical points at
\begin{equation}\label{eq:fps}
\lambda_\mathrm{UV} \equiv 0 \qquad \mathrm{and} \qquad \lambda_\mathrm{IR} \equiv \pm\log(2+\sqrt{5}) \,,
\end{equation}
corresponding to AdS$_4$ solutions with AdS radii $L_\mathrm{UV}=  \frac{1}{\sqrt{2}g}$ (the maximally supersymmetric vacuum) and $L_\mathrm{IR} =\sqrt{\frac{3}{7}}L_\mathrm{UV} $, respectively.  In order to facilitate comparison to results which have previously appeared in the literature, we will use slightly different units in this section than in section \ref{sec:SU4}, using units such that $g=1$. 

\begin{figure}
\centering
\includegraphics[scale=0.38]{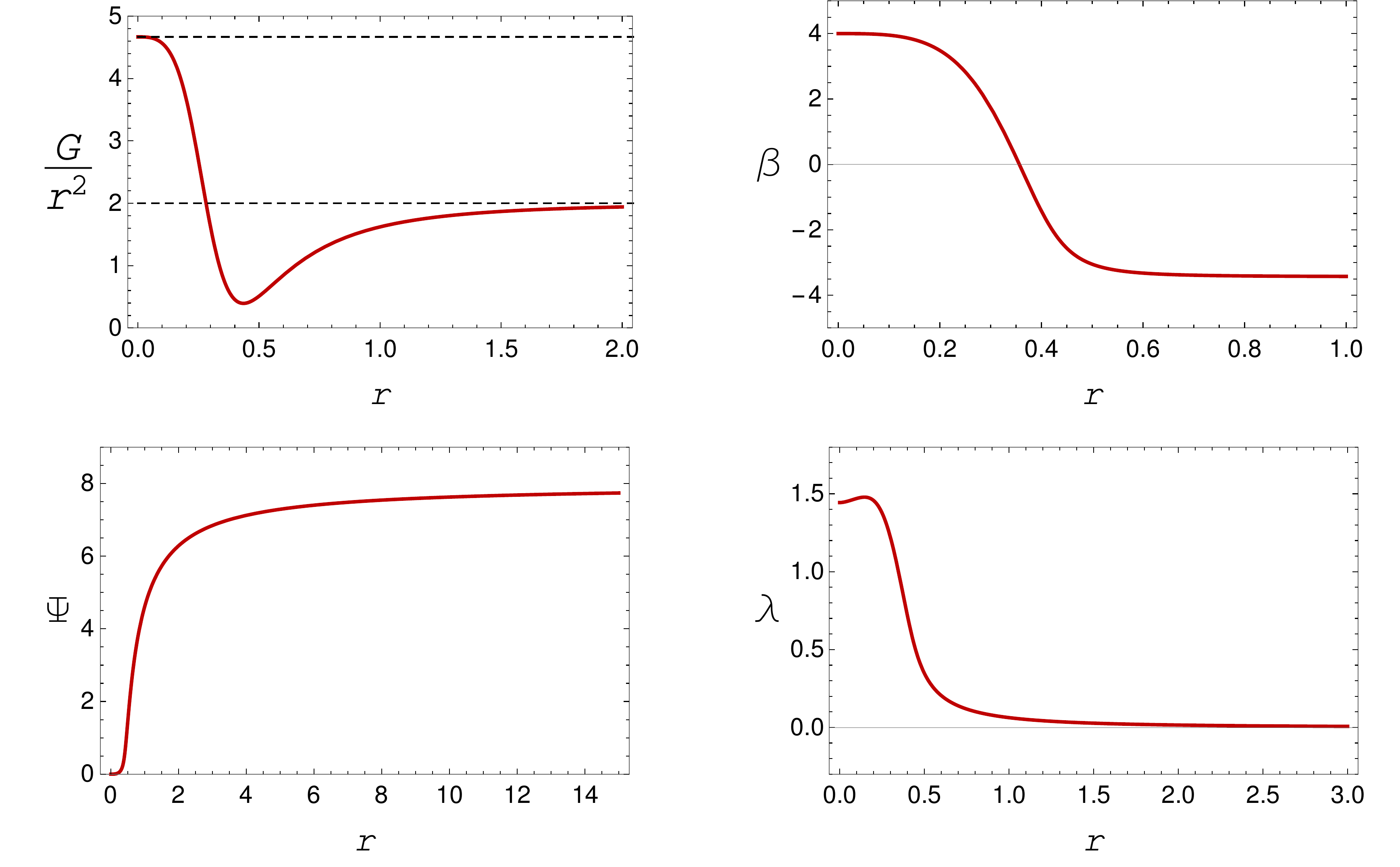}
\caption{The ``Massive Boson" background. The dashed lines in the plot of $G/r^2$ are at 14/3 and 2, indicating the values obtained in the IR and UV AdS$_4$ fixed points respectively. The ratio of the speed of light in the UV CFT compared to that of the IR theory is $n=26.900$, and the non-vanishing scalar fall-off is $\frac{\lambda_2^{1/2}}{\Psi_{\mathrm{UV}} } \approx 0.0308$.
\label{fig:typeI}}
\end{figure}

\begin{figure}
\centering
\includegraphics[scale=0.38]{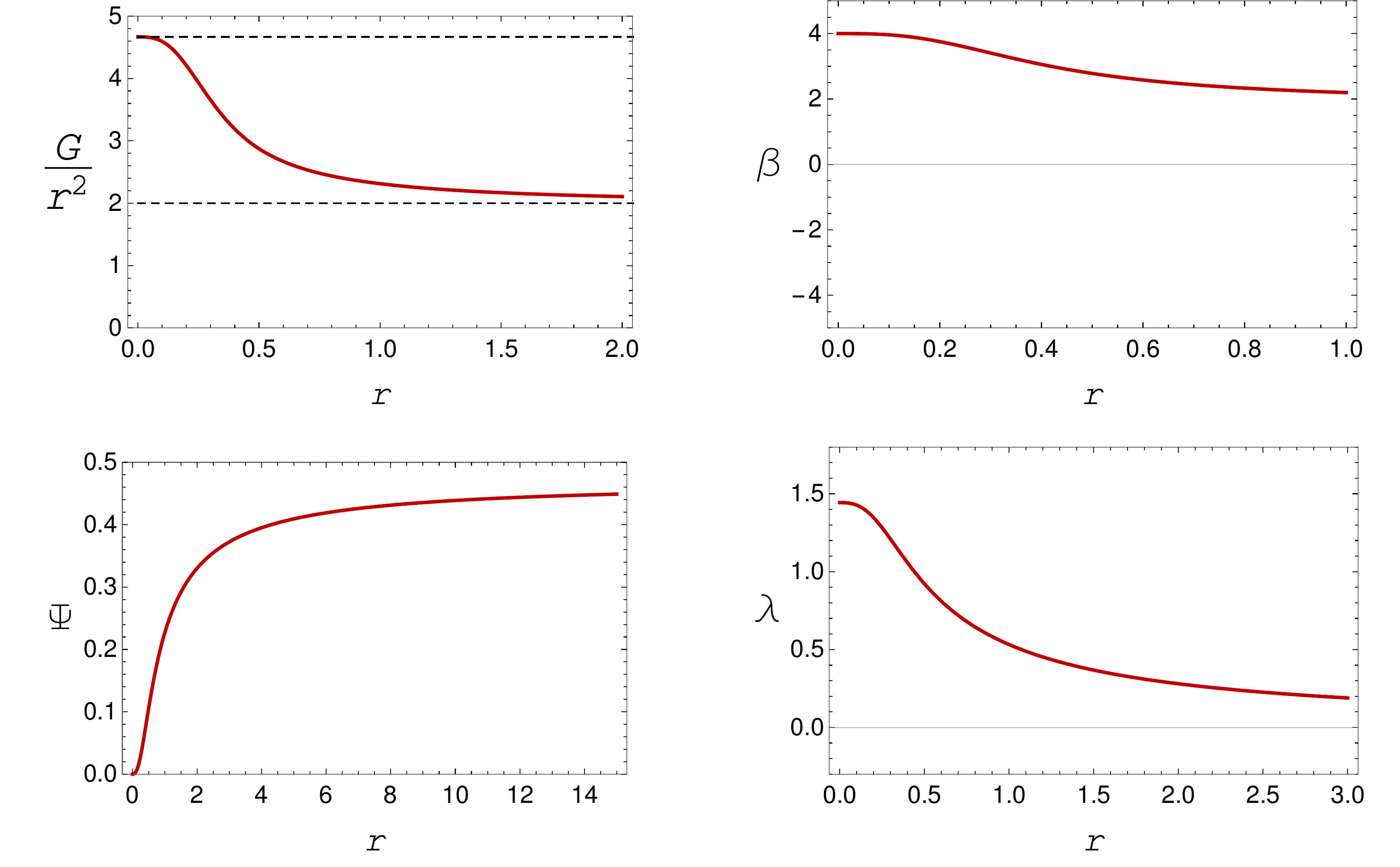}
\caption{The ``Massive Fermion" background. The dashed lines in the plot of $G/r^2$ are at 14/3 and 2, indicating the values obtained in the IR and UV AdS$_4$ fixed points respectively. This geometry is characterized by $n=1.861$ and $\frac{\lambda_1 }{\Psi_{\mathrm{UV}}} \approx 1.227$.
\label{fig:typeII}}
\end{figure}

Solutions to the equations of motion coming from (\ref{eq:Lbose}) will provide the classical backgrounds we wish to probe. Domain wall solutions in this truncation can again be described by a simple radial ansatz
\begin{equation}\label{eq:DWansatz2}
\dd s^2 = -G(r)e^{-\beta(r)}\dd t^2 + \frac{\dd r^2}{G(r)}+r^2\dd\vec{x}^2, \qquad \mathcal{A} = \Psi(r) \, dt, \qquad \mathrm{and} \qquad \lambda= \lambda(r) \,,
\end{equation}
 and a non-trivial bulk profile for $\lambda$ vanishes near the AdS boundary like $\lambda(r\to\infty)\sim \lambda_1/r + \lambda_2/r^2$. To determine what boundary conditions are interesting, we must consider the dimensionality of the operator dual to $\lambda$. As discussed previously \cite{Bobev:2011rv, DeWolfe:2015kma}, $\lambda$ is in fact dual to a linear combination of a fermion and a boson bilinear (where each bilinear also includes monopole operators). Bosonic bilinears have $\Delta=1$ while fermionic bilinears have $\Delta=2$, hence their sources are proportional to $\lambda_2$ and $\lambda_1$, respectively. This has the consequence that any solution with $\lambda$ turned on necessarily leads to \textit{explicit} symmetry breaking, being dual to ABJM theory deformed by a bilinear charged under the global U(1). With $\lambda_1 \neq 0$ we source the fermion bilinear, with $\lambda_2 \neq 0$ we source the boson bilinear. In \cite{Bobev:2011rv} two solutions corresponding to $\lambda_1=0$ and $\lambda_
2=0$ were constructed; they were further explored in the context of fermion response in \cite{DeWolfe:2015kma}. These are the solutions we will study here. Since each of these solutions sources a mass term for either a composite boson or fermion field, we will refer to them as the ``Massive Boson" and the ``Massive Fermion"  background, respectively. The solutions are shown in figures \ref{fig:typeI} and \ref{fig:typeII}. Note that the different choice of units for the SUGRA gauge coupling $g$ in this section relative to section \ref{sec:SU4} is visible in the difference in the asymptotic value of the metric function $G/r^2 \to 1/L_{\mathrm{UV}}^2$ as $r\to \infty$.

\subsection{The Fermionic Sector}

We now wish to derive the SUGRA Dirac equations in the $SO(3)\times SO(3)$ domain wall backgrounds. We first isolate a sector of spin-1/2 fermions that do not mix with the gravitini. The $SO(3) \times SO(3)$ group is embedded in the $SO(6) \simeq SU(4)$ group of the ABJM decomposition in the natural way. Under  $SO(8) \to SU(4) \times U(1)_b \to SO(3) \times SO(3) \times U(1)$  the gravitini transform as
\begin{eqnarray}\label{eq:SU4inSO32}
{\bf 8}_{\rm s}  \to {\bf 6}_0 \oplus {\bf 1}_2 \oplus {\bf 1}_{-2} \to ({\bf 3}, {\bf 1})_0 \oplus ({\bf 1}, {\bf 3})_0 \oplus ({\bf 1}, {\bf 1})_2 \oplus ({\bf 1}, {\bf 1})_{-2}\,,\ \label{eDec}
\end{eqnarray}
and thus we can avoid mixing in the $SO(3) \times SO(3)$-invariant backgrounds as long as we study fermions in representations other than these. The spin-1/2 fields are contained in the ${\bf 56}_{\rm s}$ of $SO(8)$, which decomposes as
\begin{eqnarray}\nonumber
{\bf 56}_{\rm s} &\to& {\bf 15}_2 \oplus {\bf 15}_{-2} \oplus {\bf 10}_0 \oplus {\bf \overline{10}}_0 \oplus {\bf 6}_0 \\ 
\label{56Decomp}
&\to&   ({\bf 3}, {\bf 3})_2 \oplus ({\bf 3}, {\bf 1})_2  \oplus ({\bf 1}, {\bf 3})_2
\oplus ({\bf 3}, {\bf 3})_{-2} \oplus ({\bf 3}, {\bf 1})_{-2}  \oplus ({\bf 1}, {\bf 3})_{-2}
 \oplus \label{fsDec} \\ &&  2 ({\bf 3}, {\bf 3})_0  \oplus 2 ({\bf 1}, {\bf 1})_0 \oplus ({\bf 3}, {\bf 1})_0  \oplus ({\bf 1}, {\bf 3})_0 \,.
\nonumber
\end{eqnarray}
We see there are four fermions in the $({\bf 3}, {\bf 3})$ representation of $SO(3) \times SO(3)$---a charged fermion, a neutral fermion and their conjugates---that cannot mix with the gravitini. Group theory does not prevent them from mixing with each other, and generically they do.  The different $U(1)$ charges of the fermions in the $({\bf 3}, {\bf 3})$ representations are no obstacle to this mixing because the $U(1)$ symmetry is broken by a non-trivial profile for the charged $\lambda$ in our  backgrounds. Moreover, the fact that $({\bf 3}, {\bf 3})$ is a real representation means mixing between the spinors and their conjugates is possible, meaning the Majorana coupling of \eno{MajoranaTerm} can exist.

To derive the explicit Dirac equations, we evaluate the scalar tensors in the fermionic Lagrangian (\ref{FermiLagrangian}) in the $SO(3)\times SO(3)$ truncation. As anticipated, we find mixing between sets of four fermions, corresponding to the four copies of $({\bf 3}, {\bf 3})$. We focus on only one of these sets, say \{$\chi_{467}, \chi_{538}, \chi_{418}, \chi_{428}$\}, since the other sets are related through group theory. The fermions can be assembled into complex combinations that are charge eigenstates,
\begin{eqnarray}
\label{ChargeBasis}
\chi_2=\chi_{428}+i\chi_{418}, \quad \bar\chi_{2}=\chi_{428}-i\chi_{418},& \quad \chi_0=\chi_{467}+i\chi_{538}, \quad \bar\chi_{0}=\chi_{467}-i\chi_{538} .
\label{eq:ChargeBasis}
\end{eqnarray}
The  $\chi_2$ and $\chi_0$ modes have $U(1)$ charges 2 and 0, respectively, and the barred spinors, being charge conjugates of the un-barred ones, have opposite charge. The Dirac equations for these fermions take the form
\begin{equation}\label{eq:DEQ}
\Big(i\Gamma^\mu\nabla_\mu\,{\bf 1}+ {\bf S}\Big)\vec{\chi}=0 \,,
\end{equation}
where {\bf 1} is a $4\times4$ identity matrix, $\vec{\chi} \equiv \{\chi_2, \bar\chi_2, \chi_0, \bar\chi_0\}$, and ${\bf S}$ is a mixing matrix with contributions from gauge, Pauli, and mass type couplings, whose explicit form is
\begin{equation}\label{eq:mixingMatrix}
\left(
\begin{array}{cccc}
 -\frac{1}{4} \slashed{\mathcal{A}} (\cosh 2 \lambda+3) & 0 & \Gamma_5 \sinh \lambda & - \sinh \lambda \\
 0 & \frac{1}{4} \slashed{\mathcal{A}} (\cosh 2 \lambda +3) & - \sinh \lambda & -\Gamma_5 \sinh \lambda \\
 -\Gamma_5 \sinh \lambda & - \sinh \lambda & \frac{i}{2 \sqrt{2}}\slashed{\mathcal{F} }& \frac{1}{2} \left(\slashed{\mathcal{A}}-\sqrt{2}\right) \Gamma_5 \sinh ^2 \lambda \\
 - \sinh \lambda & \Gamma_5 \sinh \lambda & \frac{1}{2} \left(\slashed{\mathcal{A}}+\sqrt{2}\right) \Gamma_5 \sinh ^2 \lambda & -\frac{i}{2 \sqrt{2}}\slashed{\mathcal{F}} \\
\end{array}
\right) .
\end{equation}
Here $\slashed{\mathcal{A}}\equiv\Gamma^{\mu}\mathcal{A}_\mu$, and $\slashed{\mathcal{F}}\equiv \Gamma^{\mu\nu}\mathcal{F}_{\mu\nu}$. This mixing matrix cannot be reduced into smaller blocks, and so we are obliged to solve a coupled system of linear differential equations.

Before solving these Dirac equations numerically, it is instructive to summarize the types of couplings the mixing matrix gives rise to, and the qualitative effects of these couplings on the fermionic spectrum. We will use the same projectors (\ref{Projectors}) as in the previous section to label the four spinor components as $\chi_{\alpha \pm}$ with $\alpha=1,2$. Writing out the Dirac equations (\ref{eq:DEQ}) at the level of the spinor components, it is easy to see that they split into two independent sets. One set couples together the $\alpha=1$ components of $\chi_2$ and $\chi_0$ with the $\alpha=2$ components of $\bar \chi_2$ and $\bar \chi_0$. The other set of equations is identical but with $\alpha=1 \leftrightarrow \alpha=2$ and $k \rightarrow -k$.

\begin{figure}
 \centering
 \includegraphics[scale=0.42]{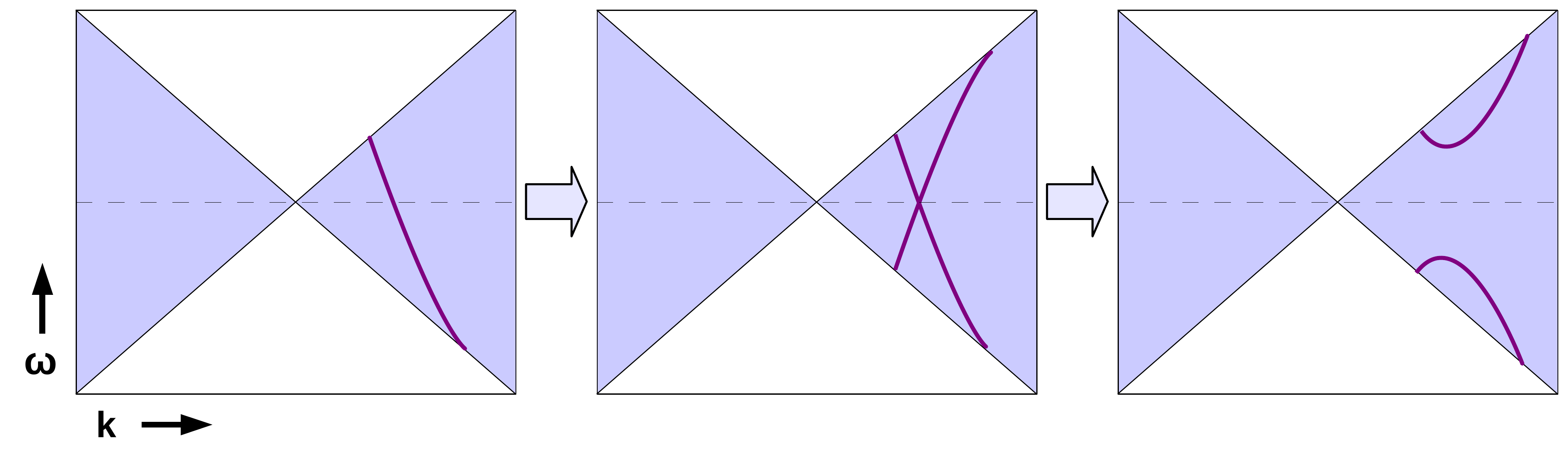}
 \caption{An illustration of the level repulsion induced by the chiral Majorana coupling in the $(\omega,k)$-plane. \textit{Left}: Without a Majorana coupling, (one of the two spinor components $\alpha$ of) a fermion operator will generically display lines of normal modes (purple) crossing the dashed $\omega=0$ line, leading to a Fermi surface singularity. \textit{Center}: Looking at the conjugate fermion, and switching to the other spinor component, gives an identical normal mode line flipped across $\omega=0$. \textit{Right}: Turning on the chiral Majorana coupling mixes these two energy bands, causing them to repel.}
 \label{fig:levelRep}
\end{figure}

This coupling is a generalization of the ``Majorana coupling" discussed by \cite{Faulkner:2009am} to involve more than one spinor field. As we described in the introduction, \cite{Faulkner:2009am} noted that such a coupling effectively forbids the existence of a holographic Fermi surface. This can be understood as a consequence of level repulsion. Imagine that in the absence of such a Majorana coupling, the $\alpha=1$ component of $\chi_2$ has a band of normal modes that crosses $\omega=0$ and at a non-zero $k=k_F$; this crossing is interpreted as a Fermi surface (left part of figure \ref{fig:levelRep}). Because $\bar\chi_2$ is the charge conjugate of $\chi_2$, it will have a similar normal mode band but with $(\omega, k) \rightarrow (-\omega, -k)$, thus it crosses $\omega=0$  at $k=-k_F$. Moreover, the spectrum of the $\alpha=1$ components is related to that of $\alpha=2$ components by $k\rightarrow -k$, as a consequence of the background rotational symmetry. Taken together, 
this means 
that the $\alpha=2$ component of $\bar\chi_2$ has a normal mode band that is related to that of the $\alpha=1$ component of $\chi_2$ by a reflection across $\omega=0$ (center of figure \ref{fig:levelRep}). In particular, these two bands will cross at $(\omega,k)=(0,k_F)$. Finally then, turning on the Majorana coupling between them will cause level repulsion at this crossing point, gapping out the Fermi surface (right part of figure \ref{fig:levelRep}). This mixing between components can be thought of as analogous to the Bogoliubov transformation mixing particles and holes in BCS theory. This prediction will be confirmed in our numerical results in the next subsection.

In field theory terms, the Majorana coupling in supergravity corresponds to the existence of a three-point function which is schematically of the form $\langle {\cal O}_\lambda {\cal O}_\chi {\cal O}_\chi \rangle$ among the operator ${\cal O}_\lambda$ dual to the active scalar $\lambda$ and the fermion.  This three-point function is visible in the vacuum state of the dual field theory, and its strength controls how strong the gapping of the Fermi surface will be.  It would be interesting to try to develop a more model-independent, field theoretic account of how similar three-point functions control the size of a superconducting gap.

Additionally, a comparison to the results of \cite{DeWolfe:2015kma} will be helpful. There, fermion response for the same quartet of SUGRA fermions in the same $SO(3)\times SO(3)$ domain walls were considered, but the chiral parts of the Majorana couplings in the fermion Lagrangian (\ref{FermiLagrangian}) were neglected. As explained above, the $\Gamma_5$ matrices in the Majorana couplings are directly responsible for the coupling between spinor components with different $\alpha$; without them the Majorana terms couple only $\alpha = 1$ to $\alpha = 1$ components. Hence, by neglecting these couplings the gapping mechanism described above is no longer present; level repulsion still occurs among the mixed fermions, but it is no longer guaranteed to be localized at $\omega=0$ where it can create a gap. Thus, when turning on the ``non-chiral" Majorana couplings in \cite{DeWolfe:2015kma}, some Fermi surface singularities were lifted, while others remained.

\subsection{Fermion Response}

We now proceed to solve our set of coupled Dirac equations. Many steps are identical to those described in section \ref{sec:SU4} and will not be repeated. The one new ingredient in this system is the mixing between different fermions through the matrix ${\bf S}$. As a consequence of this mixing, when sourcing any of the coupled fermions, there will generically be a response in all four of them. This gives rise to a matrix of Green's functions, schematically
\begin{equation}\label{eq:GRdef}
 \mathcal{G}_R^{ij} =   {\delta \langle \mathcal{O}^j\rangle \over \delta J^i} \Big\vert_{J^k = 0} \,,
\end{equation}
where $i,j \in \{1,2,3,4\}$ label the four coupled fermions, $J$ and $\langle \mathcal{O}\rangle$ denote sources for and responses of the dual operators, respectively, and $J^k=0$ implies that all sources except $J^i$ are zero. The computation of this matrix, including the implementation of correct boundary conditions, requires some care; this is described in detail in \cite{Kaminski:2009dh, Ammon:2010pg} and is implemented in a very similar system in \cite{DeWolfe:2015kma}. We refer the thorough reader to those references, and proceed directly to a discussion of our results.

First of all, as was the case in section \ref{sec:SU4}, the IR geometry controls important aspects of these Green's functions. For timelike IR momenta, as defined in (\ref{eq:IRmom}), infalling boundary conditions are imposed. These boundary conditions are complex, which can lead to quasinormal mode solutions. In the dual gauge theory these correspond to excitations with finite lifetimes. In contrast, for spacelike IR momenta one instead imposes regular boundary conditions in the IR. This is a purely real boundary condition, and may give rise to normal mode solutions in the bulk. Such solutions could correspond holographically to stable fermionic excitations in the boundary field theory. 

\begin{figure}
 \centering
 \includegraphics[scale=0.5]{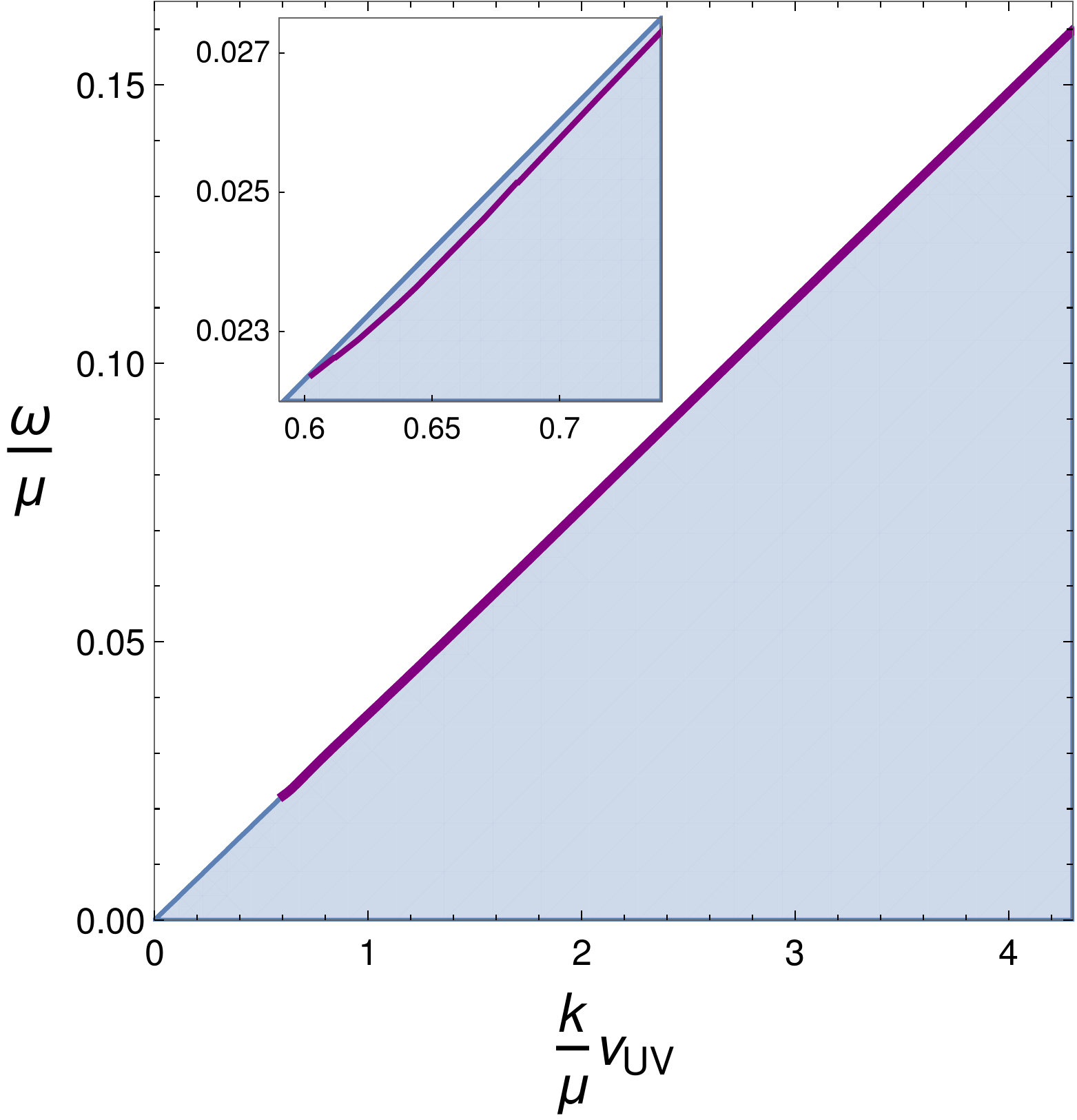}
 \caption{The band structure of fermion normal modes in the Massive Boson (type 1) background. The normal mode is shown  in purple. The inset zooms in on the beginning of this band, emphasizing that it very nearly coincides with the edge of the IR lightcone.}
 \label{fig:MB_NM}
\end{figure}

If the fermion spectral weight is non-vanishing at zero frequency, it means that there are gapless fermionic modes in the dual phase. In  \cite{DeWolfe:2015kma}, an interesting prediction of the ``nearly top-down" model was that spectral weight appeared as a band of delta functions passing through a Fermi surface singularity at $\omega=0$ and $k=k_F$. However, as anticipated previously by the chiral Majorana coupling-induced level repulsion argument, we find that the true top-down system admits very few normal modes at all. In the Massive Fermion background we find none, while in the Massive Boson background there is a line of normal modes very close to the lightcone, as seen in figure \ref{fig:MB_NM}. These lines sit very close to the lightcone edge, and go on to quite large $k$ and $\omega$.

\begin{figure}
 \centering
 \includegraphics[scale=0.42]{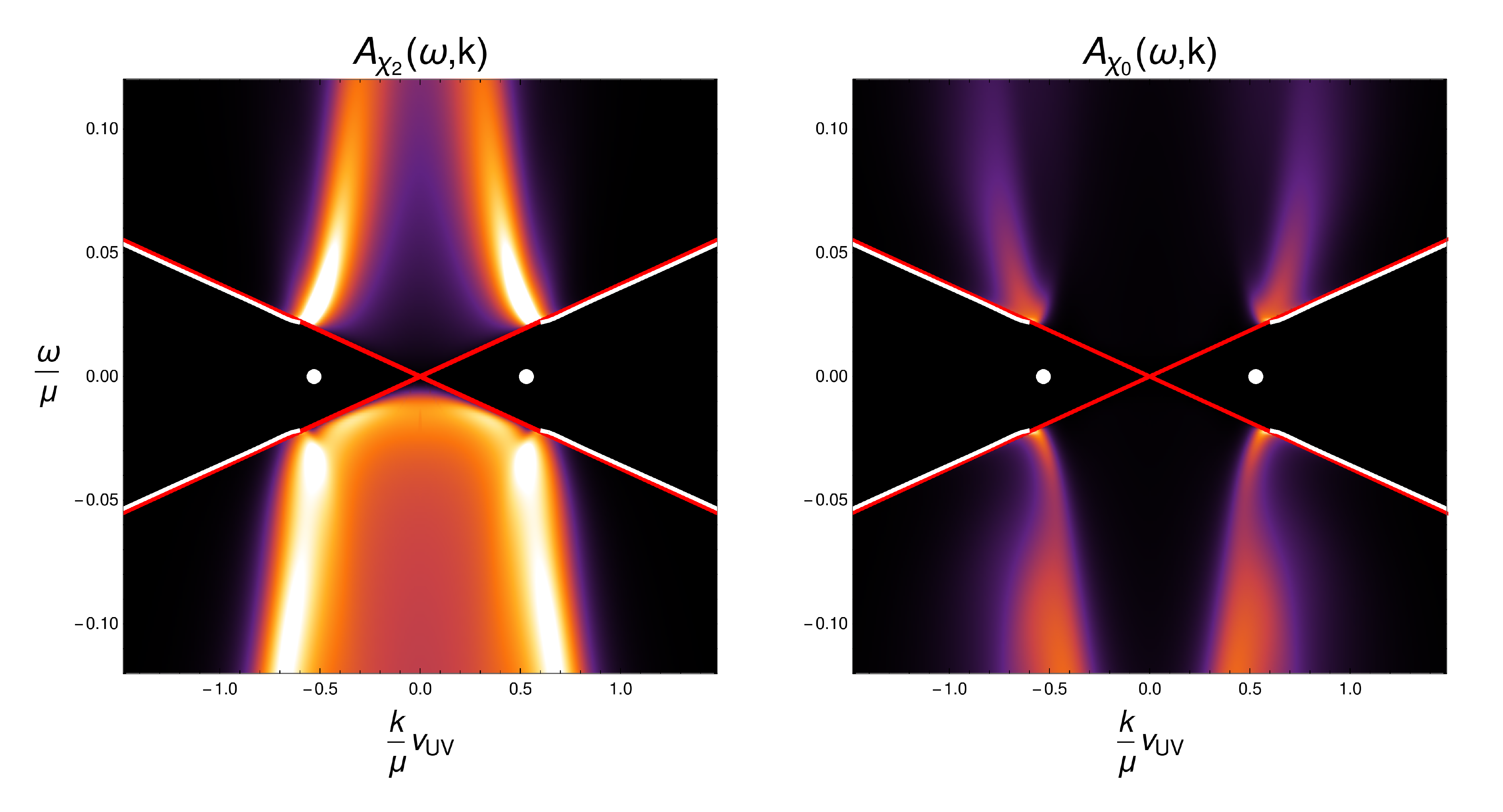}
 \caption{The spectrum in the Massive Boson background. The red and blue lines mark the IR and UV lightcones, respectively, and the white lines show the location of the line of normal modes, corresponding to a line of delta function peaks in the spectral weight. The white dots at $\omega=0$ show the Fermi momentum in the normal phase.}
 \label{fig:MBspectrum}
\end{figure}

\begin{figure}
 \centering
 \includegraphics[scale=0.42]{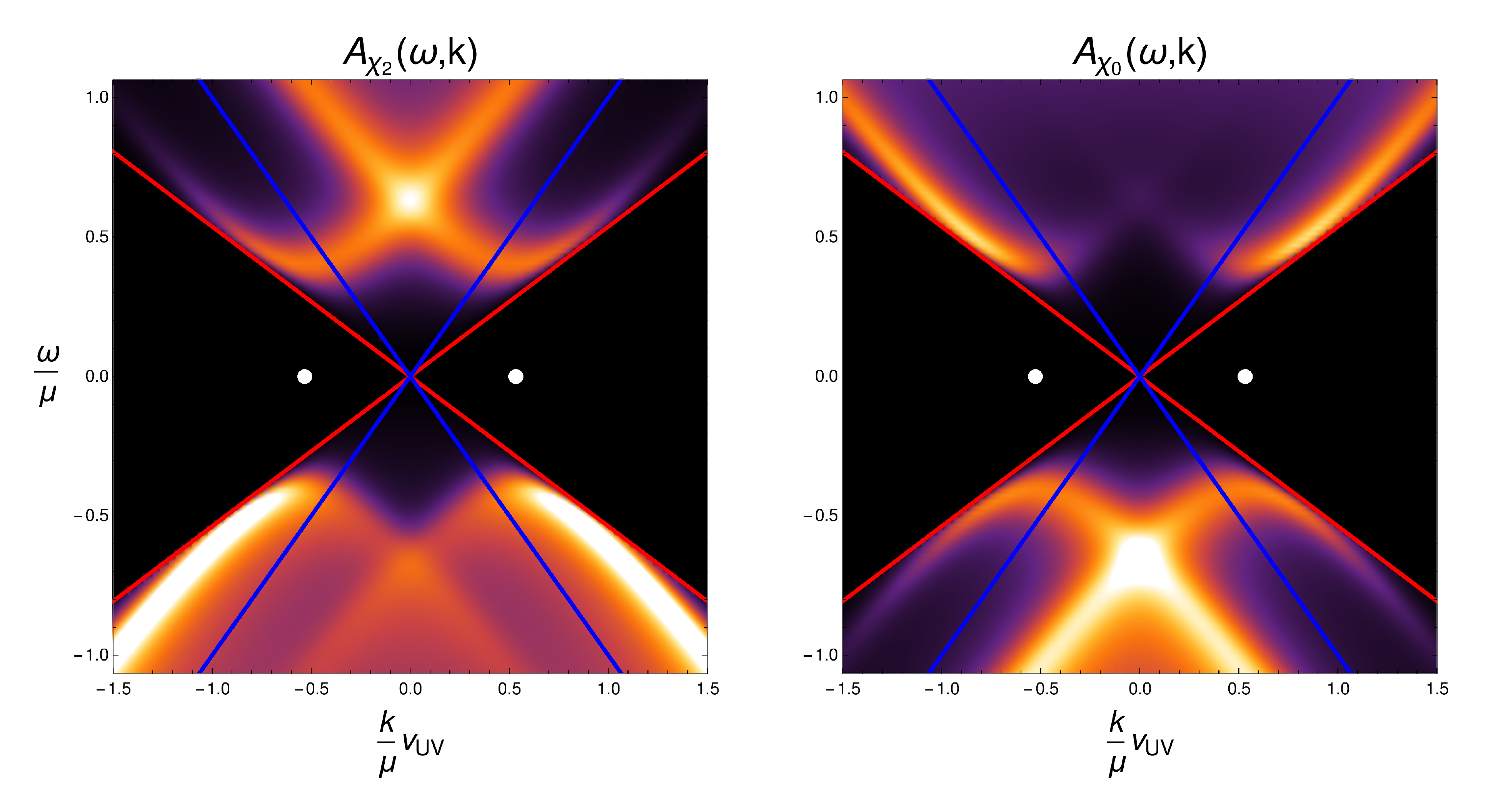}
 \caption{The spectrum in the Massive Fermion background. The red and blue lines mark the IR and UV lightcones, respectively; for spacelike IR momenta the spectral weight is zero everywhere. The white dots at $\omega=0$ show the Fermi momentum in the normal phase.}
 \label{fig:MFspectrum}
\end{figure}

To get a more detailed picture of the spectrum, we plot the spectral functions as defined in (\ref{eq:spectralFunc}) for the fermions of the $({\bf 3},{\bf 3})$ in the Massive Boson and Massive Fermion backgrounds in figures \ref{fig:MBspectrum} and \ref{fig:MFspectrum}, respectively. As in section \ref{sec:SU4}, we observe arcing spectral weight inside the IR lightcone presumably due to the presence of bulk fermion quasinormal modes. Particularly in figure \ref{fig:MFspectrum} we again observe a crossing of the arcs coming from different spinor components. The mixing of charged and neutral fermions is seen in the transfer of spectral weight between arcs as one follows them while varying $k$ (this is most clearly seen in the massive fermion background). Note that the massive boson normal modes are some ($k$- and $\omega$-dependent) linear combination of the charged and neutral fermions, hence they are drawn in both plots. Importantly, in both holographic phases the spectral weight is only non-zero away from $\omega=0$, and in nearly every case there is a fairly pronounced gap in the spectral function. Hence we find no sign of a 
Fermi surface in the fermion correlation functions in these holographic states. The figures show the spectral functions of $\chi_2$ and $\chi_0$; the spectral functions of their charge conjugate modes are identical but with $\omega \rightarrow -\omega$, as discussed above.

\subsection{Field Theory Operator Matching}
Once more exploiting our top-down framework, we can write down exactly which operators in ABJM theory are dual to the quartet of fermions under study. Unlike the $SU(4)^-$ case, here the symmetry structure aligns nicely with the ABJM decomposition of $SO(8)$ described above (\ref{eq:SU4inSO32}),
with $SO(3) \times SO(3)$ embedded in $SO(6) \simeq SU(4)$ in the natural way. As a result, we can use ABJM operator language directly. This was worked out in \cite{DeWolfe:2015kma}, and for details we refer the reader there.  Here, we simply quote the results:
\begin{eqnarray}
\chi_2  \quad &\leftrightarrow& \quad 
 \left( Y^1 \psi_2 - Y^2 \psi_1 + Y^3 \psi_4 - Y^4 \psi_3 \right) e^{2 \tau} \,, \label{eq:chip}\\
\bar\chi_2 \quad &\leftrightarrow&\quad 
\left( Y^\dagger_1 \psi^{\dagger 2} - Y^\dagger_2 \psi^{\dagger 1} + Y^\dagger_3 \psi^{\dagger 4} - Y^\dagger_4 \psi^{\dagger 3} \right) e^{-2\tau}\,,\\
\chi_0  \quad &\leftrightarrow& \quad 
 Y^1 \psi^{\dagger 4} + Y^4 \psi^{\dagger 1} - Y^2 \psi^{\dagger 3} - Y^3 \psi^{\dagger 2}  \,, \\
\bar\chi_0  \quad &\leftrightarrow& \quad 
Y^\dagger_1 \psi_4 + Y^\dagger_4 \psi_1 - Y^\dagger_2 \psi_3 - Y^\dagger_3 \psi_2  \,.\label{eq:chib0}
\end{eqnarray}
In this mapping, the $Y$'s are ABJM scalars, $\psi$'s are fermions, and $e^{2\tau}$ is a monopole operator which carries all of the charge under the $U(1)_b$. 

This identification of symmetries facilitates the field theory description, allowing one to interpret the dual state of matter as a phase in which a chemical potential for monopole operators has been turned on. The four Cartan chemical potentials are identified as
\begin{equation}
\label{SO3SO3Mu}
	\mu_a = \mu_b = \mu_c = \mu_d \,.
\end{equation}
This corresponds to the gauge field $A_{12}$ alone being turned on because $1, 2$ are ${\bf 8}_{\rm s}$ indices, and a triality rotation to the ${\bf 8}_{\rm v}$ basis reveals all four Cartan charges are turned on equally. The non-trivial bulk scalar signals an explicit breaking of the number density for this composite matter, by an operator of the form
\begin{equation}
\mathcal{O}_{\Delta=1}\sim Y^AY^Ae^{2\tau}\qquad \mathrm{or}\qquad  \mathcal{O}_{\Delta=2}\sim\psi^A\psi^Ae^{2\tau}\,,
\end{equation}
for the massive boson (fermion) case, respectively. Viewed in this language, our results for the massive fermion phase demonstrate a novel phase of strongly coupled matter in which there exist perfectly stable composite fermion excitations above a hard gap.

\section{Lessons for Strongly Coupled Systems}\label{sec:lessons}
One of the most striking lessons from our calculation of fermion spectral functions is that in the broken symmetry phases of ABJM matter that we study, our fermion spectral densities are always gapped. This observation merits further discussion, as it appears to manifest for different reasons in the two cases, and it is not entirely clear how generic this result might be. In an attempt to better understand the absence of Fermi surface singularities in these spectral functions, it proves useful to compare our results against several related calculations which we now describe.

\subsection{Top-down vs. Bottom-up Fermion Response}
In previous investigations of fermion spectral functions in domain wall flows \cite{Gubser:2009dt,DeWolfe:2015kma}, the authors employed non-top-down fermions in an attempt to gain intuition for how the fermionic degrees of freedom behave in the dual phases of matter. A surprising result was the presence of families of bulk fermion normal modes which collectively described ungapped bands of perfectly stable fermionic excitations in the dual field theory.

To realize these bands, it is necessary to deform our top-down system by ignoring the constraints that $D=4$ maximal gauged SUGRA places on the bulk fermion couplings. In the $SU(4)^-$ flow, for example, one can make contact with \cite{Gubser:2009dt} by setting the scalar to zero in the top-down Dirac equation (\ref{eq:DEQ20}) (so that the fermion couplings do not run), dropping the Pauli coupling, and artificially dialing the bulk fermion's charge. As explained in \cite{Gubser:2009dt}, for suitably large values of this ``probe" fermion's $U(1)$ charge, ungapped bands of normal modes appear and a holographic Fermi surface is present. To study the sensitivity of our results from the $SU(4)^-$ flow, we computed the spectral function for a number of such deformations of (\ref{eq:DEQ20}). We find that setting the scalar to zero in the Dirac equation, but otherwise leaving the magnitude of the couplings untouched, leaves the results largely unchanged; in particular, the gap remains. However, if we 
additionally tune the couplings by $\mathcal{O}(1)$ factors, for example by doubling the charge or changing the sign of the Pauli coupling, the gap will in general close. This is consistent with the results of \cite{Gubser:2009dt}, which show that the larger the fermion charge, the more bands of gapless modes are present.

In this context then, it would seem that the fermion spectral functions in the $SU(4)^-$ domain wall of section \ref{sec:SU4ferm} end up gapped for a fairly straightforward reason: SUGRA demands that in this state, the fermions in the ${\bf 20}$ carry a $U(1)$ charge that is too small to support a Fermi surface.

This stands in contrast to the gapping mechanism that appears to be at work in the $SO(3)\times SO(3)$ flow. The results of \cite{DeWolfe:2015kma} demonstrate that in this phase, the $U(1)$ charge carried by the bulk fermion is sufficient to form a holographic Fermi surface, provided that one removes the chiral Majorana couplings by hand. (Purely bottom-up fermions with the same mass and charge, also studied there, have yet more gapless bands.)  In other words, the mechanism of \cite{Faulkner:2009am}, in which the chiral Majorana couplings play the key role, makes the difference in this case between an ungapped Fermi surface and gapped behavior.  

Thus, we find that the SUGRA couplings conspire to gap out the spectral weights in all the cases we study. However, while the resulting spectral weights all have similar features, with gapped, arcing bands, the various bulk Dirac equations have qualitative differences. The Majorana coupling \eno{MajoranaTerm} acts much like a bulk version of the BCS mechanism, and can therefore be expected to lead to the observed gaps in spectral weights. Yet the fermion spectral functions in the $SU(4)^-$ background emphasize that a gap may appear without this coupling. The precise interpretation of these different gapping mechanisms in terms of the physics in the boundary field theory deserves further investigation. Furthermore, it would clearly be interesting to study fermionic spectral weights in other top-down realizations of zero-temperature symmetry-broken states, in order to find out how general the formation of a gap really is.

\subsection{Extremal AdSRN and Effects of Broken Symmetry}

A complimentary line of insight is directed along comparisons between the spectral functions in our domain wall flows and those in states of {\it un}broken $U(1)$ symmetry. Such states are readily accessible to our decoupled fermions. They are solutions to the bosonic sectors described by (\ref{eq:Lag}) and (\ref{eq:Lbose}), but with the scalar set to zero. These backgrounds are the familiar $AdS_4$ Reissner-Nordstr\"om (AdSRN) solution, and its extremal limit is holographically dual to a distinct zero temperature finite density phase. 

Although the form of the AdSRN solution is basically the same in both the $SU(4)^-$ and $SO(3)\times SO(3)$ truncations, their holographic interpretation is slightly different because the $U(1)$ gauge fields under which the black holes are charged and the associated chemical potentials are embedded differently into $SO(8)$, as is spelled out in (\ref{SU4Mu}) and (\ref{SO3SO3Mu}). 

Nonetheless, both the fermions in the ${\bf 20}$ as well as those of the $({\bf 3},{\bf 3})$ behave similarly in their respective AdSRN backgrounds. Importantly, both systems display Fermi surface singularities in their dual fermion spectral functions. For the fermions in the $({\bf 3},{\bf 3})$, the charged modes decouple from their neutral counterparts when the scalars vanish, and unsurprisingly it is the spectral function for the charged operators that exhibits a Fermi surface.  It is perhaps helpful to emphasize that these results (unlike the previous subsection) are truly top-down. Both the AdSRN backgrounds and the spin-1/2 Dirac equations can be embedded in the maximal gauged SUGRA theory.

The results of our present work show that breaking the $U(1)$ either spontaneously or explicitly destroys this Fermi surface and gaps the corresponding spectral functions. Notably, the new state with broken symmetry appears to ``remember" the location of the Fermi surface that was present in the unbroken phase. This is demonstrated by the arcing spectral weights in figure \ref{fig:spec} (right plot) and in figures \ref{fig:MBspectrum}  and \ref{fig:MFspectrum}, which bend towards $\omega/\mu = 0$, and achieve their closest approach at some finite momentum $k^{\star}\,v_{\mathrm{UV}}/\mu$.  Computation of the fermion response in the unbroken phase reveals a Fermi surface singularity at $k_F\,v_{\mathrm{UV}}/\mu\approx 0.25$ for fermions in the ${\bf 20}$ of $SU(4)^-\subset SO(8)$, and at $k_F\,v_{\mathrm{UV}}/\mu \approx 0.53$ for the fermions in the $({\bf3},{\bf 3})$ of $SO(3)\times SO(3)\subset SO(8)$.\footnote{Note that due to the different units employed in sections \ref{sec:SU4} and \ref{sec:SO3xSO3},  
care should be taken in comparing the Fermi momenta between the two phases.} From the figures, one finds that indeed $k^{\star}/k_F \sim 1$.

\begin{figure}
 \centering
 \includegraphics[scale=0.39]{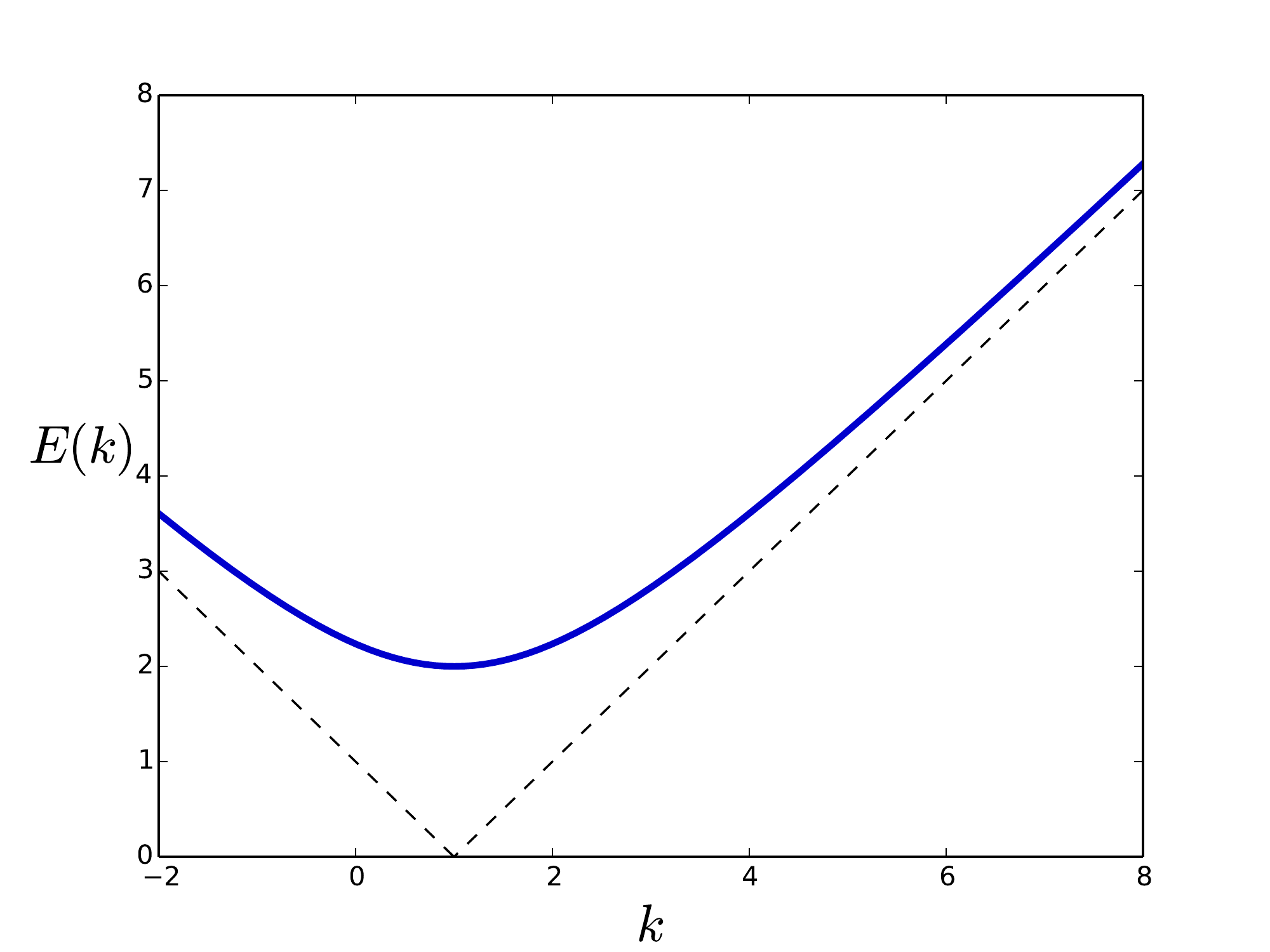}
 \includegraphics[scale=0.22]{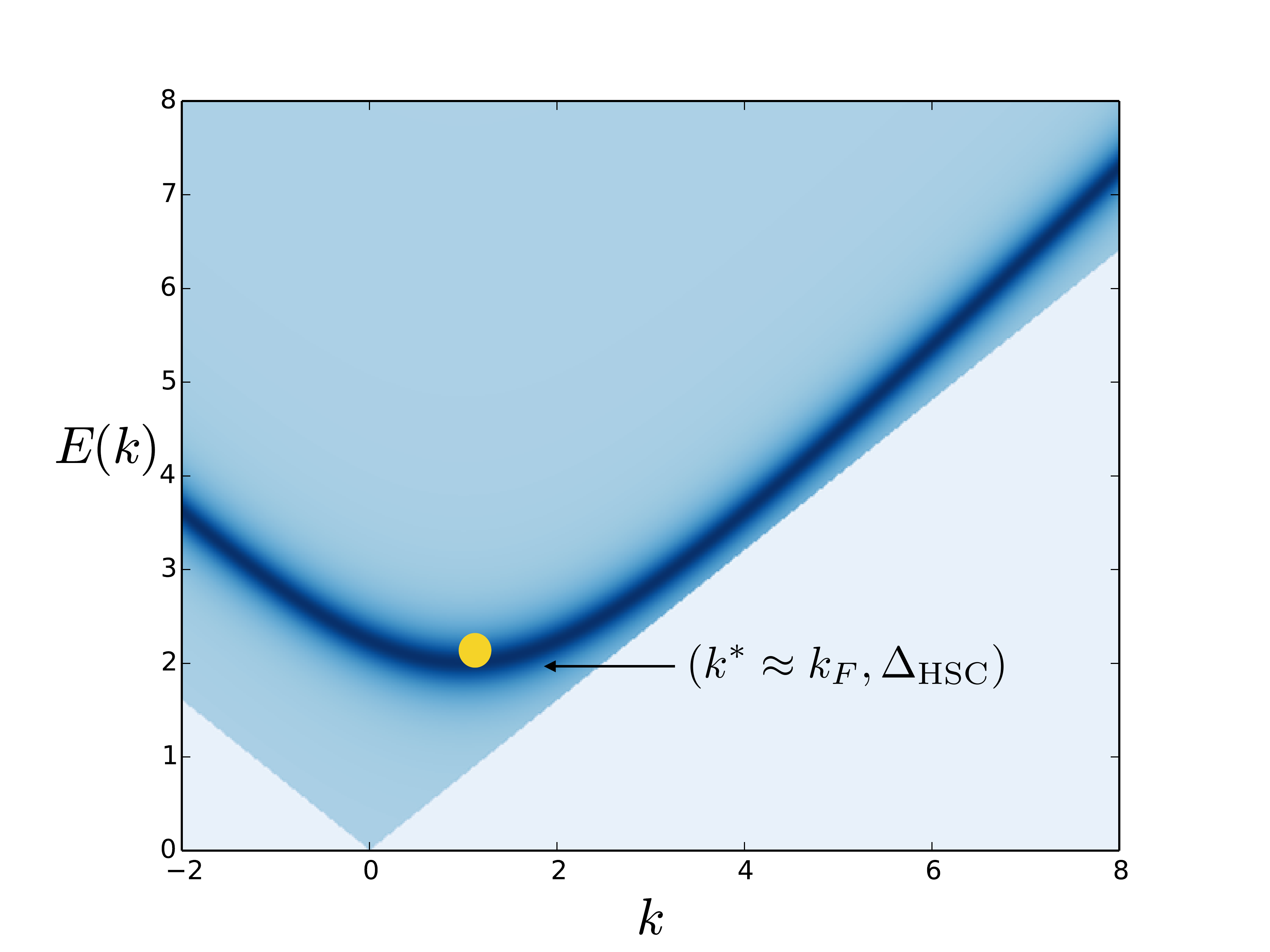}
 \caption{Illustration of gapped fermionic excitations in BCS theory and holography. In the left panel, the BCS dispersion relation in the superconducting (normal) phase is plotted in blue (dashed black). The parameters are arbitrarily chosen such that $v_F = k_F = 1$ and $|\Delta| =2$. In the holographic fermion spectral function (cartoon, right), the boundaries of the IR lightcone determines the stability of the fermionic excitations, but the gapping is qualitatively similar.}
 \label{fig:gapdisp}
\end{figure}

It is interesting to compare this to the gapping that occurs in the fermionic excitation spectrum  of the standard BCS theory. In the normal phase of a superconductor, particles and holes have an approximately linear dispersion about the Fermi surface at $k=k_F$. Thus, in a rotationally invariant system, $\epsilon(k)\approx v_F(k-k_F)$ with $v_F$ the Fermi velocity. As the superconductor is cooled into the superconducting phase, Cooper pairs condense and the mean field BCS Hamiltonian can be rediagonalized via a Bogoliubov transformation that mixes particles and holes. These new Bogoliubov modes describe the fermionic excitations in the superconducting phase, and have a dispersion relation of the form
\begin{equation}
E(k) = \sqrt{\epsilon(k)^2 +|\Delta|^2}\approx \sqrt{v_F^2(k-k_F)^2+|\Delta|^2},
\end{equation}
which is plotted in the left panel of figure \ref{fig:gapdisp}.

In the right panel of the same figure, a sketch comparing some related features in figures \ref{fig:spec}, \ref{fig:MBspectrum},  and \ref{fig:MFspectrum} is shown. The cartoon emphasizes the arcs in the spectral weight for fermionic excitations, whose minima at $k^\star \approx k_F$ define a gap that is present in the holographic results.\footnote{While the spectral functions we compute have a ``soft gap" at $k=0$ in the sense that the spectral weight vanishes as a power law in $\omega$ (see \textit{e.g.} figure \ref{fig:SU4_k0}), the majority of the spectral weight is concentrated into these (gapped) arcs.} Also depicted is the qualitative effect of the IR critical point, which opens a window of stability for any excitations that may be present in the kinematic region defined by the exterior of the IR lightcone. In the illustration there are no such stable excitations,  but such excitations do appear in the spectrum of fluctuations in the Massive Boson background (figure \ref{fig:MBspectrum}).

It is worth noting that the peaks of the various spectral weight arcs we observe are in general not sharpest at $k = k^\star$, where the gapped  excitation achieves its lowest energy; this can be seen particularly well in the right plot of figure \ref{fig:SU4_k0}. Instead, the peak representing the gapped excitation typically sharpens further as it nears the IR lightcone. This behavior is natural from the perspective of the dual field theory, where the presence of the IR lightcone can be interpreted as the existence of a kinematic regime in which interactions mediating decays of the fermionic excitations are forbidden.

The holographic spectral densities suggest a suitable (but somewhat rough) estimate for the size of the gap in the holographic broken symmetry phases, $\Delta_{\mathrm{HSC}}$. In the examples shown in this work, the value of the excitation energy at $k^\star$ is always close to the boundary provided by the IR lightcone. Thus we can write
\begin{equation}
|\Delta|_{\mathrm{HSC}}\equiv E(k^\star)\approx E(k_F)  \sim v_{\mathrm{IR}} k_F 
\end{equation}
where $v_{\mathrm{IR}}$ is the effective speed of light in the IR theory, and $k_F$ is the value of the Fermi momentum in the symmetry unbroken phase dual to the extremal AdSRN solution. This type of estimate, while fairly accurate in our top-down realizations, will generally not be obeyed in an arbitrary bottom-up construction where one is free to tune the different couplings. Again it would be interesting to study other similar top-down embeddings in order to investigate whether this is a standard feature of such states.

\subsection{Stability in Supergravity and Zero Temperature Response}
The utility of the fermionic spectral functions is contingent on their ability to quantify and elucidate properties of interesting strongly correlated phases. While we have applied this tool to better understand how some of these phases are constructed from ABJM matter, it is also important to address  the possibilities that these zero temperature states  have to actually be realized in the phase diagram for ABJM matter at finite density. 

Fundamentally, this is a question of stability.  A useful  example is provided by the $SU(4)^-$-invariant flow of section $\ref{sec:SU4}$ and the AdSRN solution that also solves the equations of motion derived from (\ref{eq:Lag}). Very generally, both solutions holographically describe zero-temperature phases of strongly interacting ABJM matter at finite density. In both cases, the ABJM theory remains undeformed by the application of any additional sources beyond the chemical potential. Thus, it is natural to wonder which (if either) of these solutions provides the thermodynamically preferred phase for such ABJM matter at low temperatures.

Neither the $SU(4)^-$-invariant flow nor the extremal AdSRN solution preserve any of the supersymmetries of the vacuum $AdS_4$. Accordingly there is no guarantee that either solution is stable at zero temperature, and it is necessary to consider the whole spectrum of SUGRA fluctuations to hunt for instabilities. Unstable modes may, or may not, belong to the consistent truncation that results in the maximal gauged SUGRA of section \ref{sec:SUGRA}, and thus the identification of all possible instabilities is a rather involved task.

It is by now well known that extremal AdSRN solutions  exhibit a multitude of instabilities in gauged SUGRA theories. These instabilities are often diagnosed by studying the mass spectrum of supergravity fluctuations around the $AdS_2$ factor of the near horizon geometry of the extremal solution. If the fluctuation's effective mass lies below the Breitenlohner-Freedman bound \cite{Breitenlohner:1982bm, Breitenlohner:1982jf} of this IR $AdS_2$ region, an instability to the formation of a new branch of solutions with a non-trivial profile for the unstable mode is anticipated. 

In the context of the present work, this is exemplified in the ``superfluid" instability of the extremal AdSRN solution to the formation of scalar $\xi$ hair. The $SU(4)^-$-invariant flow studied in section \ref{sec:SU4} is the zero temperature endpoint of a branch of solutions which extends to finite temperatures via a series of hairy black holes which terminate at some temperature $T_c$. By comparing the thermodynamic free energy of the hairy black holes to that of the AdSRN solutions, it is straightforward to demonstrate that the solutions with $\xi$ hair are thermodynamically preferred, and that as the finite density system cools there is a second order phase transition at $T_c$ from the symmetry unbroken ``normal" phase to a broken symmetry superfluid phase with a non-vanishing condensate of the operator holographically dual to $\xi$.

Interestingly, in \cite{Bobev:2010ib} the authors demonstrate that this superfluid instability is not the end of the story at low temperatures. They show that the PW solution which characterizes the IR of the $SU(4)^-$-invariant flow is itself unstable to fluctuations of scalar modes within the gauged SUGRA, and identify the origin of these unstable modes from the eleven dimensional perspective. Consequently, the $SU(4)^-$-invariant flow cannot describe a true ground state for strongly interacting ABJM matter. 

Further instabilities in the finite-temperature generalizations of the $SU(4)^-$ flow and its AdSRN companion were identified, and the backreacted geometries corresponding to those instabilities were constructed, by \cite{Donos:2011ut} in a larger $SU(3)$-invariant truncation containing additional scalars that includes the $SU(4)^-$-invariant case as a subtruncation. These other branches of solutions are in thermodynamic competition with the branch we consider, although it is generally not known what their zero-temperature limit is.

Stability of the $SO(3)\times SO(3)$-invariant flow has been investigated in \cite{Fischbacher:2010ec}. The authors find in this case that despite lacking any supersymmetry, the IR $AdS_4$ solution is stable to scalar perturbations in the gauged SUGRA. While this stability does not automatically extend to the full flow, nor does it guarantee an absence of unstable modes in the eleven dimensional theory, it is nonetheless an interesting observation that distinguishes this flow in the context of holographic phases of matter. 

\begin{centering}
\subsubsection*{Acknowledgments}
\end{centering}
\noindent We would like to thank Daniel Dessau, Jerome Gauntlett and Chaolun Wu for useful discussions. The work of O.D.\ and O.H.\ was supported by the Department of Energy under Grant No.~DE-FG02-91-ER-40672.  O.H.\ was also supported by a Dissertation Completion Fellowship from the Graduate School at the University of Colorado Boulder. The work of S.S.G.\ was supported in part by the Department of Energy under Grant No.~DE-FG02-91ER40671.  The work of S.S.G.\ was carried out in part at the Aspen Center for Physics, supported by NSF grant PHY-1066293. The work of C.R.\ is supported by the European Research Council under the European Union's Seventh Framework Programme (FP7/2007-2013), ERC Grant agreement ADG 339140.

\appendix
\numberwithin{equation}{section}
\section{Spinor Conventions}\label{sec:spincon}
The four dimensional gamma matrices $\Gamma^a$ generate $\mathrm{Cliff}(3,1)$ and satisfy 
\begin{equation}
\left\{\Gamma^a,\Gamma^b \right\} = -2\eta^{ab} \,,
\end{equation}
with $\eta = \mathrm{diag}(-,+,+,+)$.  The frame indices $a,b$  take values  $\hat{t},\hat{r},\hat{x},\hat{y}$.  A convenient basis for these matrices (which has been used throughout this work) is provided by the decomposition
\begin{equation}\label{eq:gbasis}
\Gamma^{\hat{t}} = \sigma_1\otimes 1\qquad \Gamma^{\hat{r}} = i\sigma_3\otimes 1\qquad \Gamma^{\hat{x}} = i\sigma_2\otimes\sigma_3\qquad \Gamma^{\hat{y}} = i\sigma_2\otimes\sigma_1.
\end{equation}
This basis diagonalizes the projectors $\Pi_\alpha$ and $P_{\pm}$ which were used in the text to isolate the components of the bulk spinor which contain the source for and response of the dual field theory operator. Consequently, the spinor Green's function is also diagonal in this basis.

The chiral projectors introduced in section \ref{sec:SUGRA} are defined by
\begin{equation}
 \Gamma_5 \equiv i \Gamma^{\hat{t}} \Gamma^{\hat{x}} \Gamma^{\hat{y}} \Gamma^{\hat{r}} \,, \quad \quad P_L \equiv {1 \over 2} \left( 1 - \Gamma_5\right) \,, \quad \quad 
P_R \equiv {1 \over 2} \left( 1 + \Gamma_5\right) \,.
\end{equation}

\bibliographystyle{ssg}
\bibliography{flowsBib}
\end{document}